\documentclass{aa}  

\usepackage{graphicx}
\usepackage{txfonts}
\begin{document}

\title{Generalizing MOND to explain the missing mass in galaxy clusters}

   \subtitle{}

   \author{Alistair O. Hodson\inst{1}
          \and
          Hongsheng Zhao\inst{1}
          }

   \institute{$^1$SUPA, School of Physics and Astronomy, University of St Andrews,
             Scotland\\
              \email{aoh2@st-andrews.ac.uk}
              }

   \date{Received , ; accepted , }

 
  \abstract
   {Modified Newtonian
Dynamics (MOND) is a gravitational framework designed to explain the astronomical
observations in the Universe without the inclusion of particle dark matter.
Modified Newtonian Dynamics, in its current form, cannot explain the missing
mass in galaxy clusters without the inclusion of some extra mass, be it in
the form of neutrinos or non-luminous baryonic matter. We investigate whether
the MOND framework can be generalized to account for the missing mass in
galaxy clusters by boosting gravity in high gravitational potential regions.
We examine and review Extended MOND (EMOND), which was designed to increase
the MOND scale acceleration in high potential regions, thereby boosting the
gravity in clusters. }
   {We seek to investigate galaxy cluster mass profiles in the context of
MOND with the primary aim at explaining the
missing mass problem fully without the need for dark matter.}
   {Using the assumption that the clusters are in hydrostatic equilibrium, we can compute the dynamical mass of each cluster and compare the result to the predicted mass of the EMOND formalism. }
   {We find that EMOND has some success in fitting some clusters, but overall has issues when trying to explain the mass deficit fully. We also investigate an empirical relation to solve the cluster problem, which is found by analysing the cluster data and is based on the MOND paradigm. We discuss the limitations in the text.}
   {}


   \maketitle

\section{Introduction}

To explain dynamics in astrophysical system fully, a solution to the missing mass problem that plagues astrophysics to this day must be imposed. Commonly, the missing mass in our Universe is explained away by the existence of some non-baryonic dark matter. Although dark matter is thought by most to be the best solution to the missing mass problem, despite our best efforts to date,  the lack of direct detection has led some to explore other explanations. To do this, it is possible to think of Newtonian gravity as a limit of a more general gravity law. An example of a modified theory of gravity, designed with the absence of dark matter in mind, is Modified Newtonian Dynamics, hereafter MOND. The MOND paradigm works with the principle that Newtonian gravity breaks down in the low acceleration limit ($a<<a_{0}$ where $a_{0}$ is an acceleration constant $\approx 1.2 \times 10^{-10}\text{ms}^{-2}$). 

Originally, MOND was proposed and introduced as an empirical law in the 1983 Milgrom 1983 papers \citep{milgrom19832}, \citep{milgrom19833} and \citep{milgrom19831} by analysing galaxy rotation curves. One issue with the original MOND formulation by Milgrom, which was highlighted in \cite{mondmomentum}, was its inability to conserve momentum. This was rectified in the Bekenstein and Milgrom paper \citep{bekenstein1984} by introducing a Lagrangian formulation of MOND, AQUAL (A QUAdratic Lagrangian). Recently, Milgrom proposed an alternative formulation to the MOND paradigm called Quasi-Linear MOND \citep{milgromQUMOND}, hereafter QUMOND. This new formulation is much more desirable as the quasi-linear way in which the MOND gravity is calculated is much simpler than in AQUAL.

The MOND framework has enjoyed a vast amount of success, providing reasonable descriptions of the dynamics in galaxies. Arguably the most famous prediction of MOND is that of the baryonic Tully-Fisher relation, which shows a correlation with total enclosed baryonic mass and galactic outer rotation velocity (the flat part of the rotation curve) \citep{tullyfisher}. This correlation is explained naturally in the MOND paradigm. Also, MOND is very successful in predicting galaxy rotation curve profiles; see for example \citet{sandersreview}. A further prediction of MOND was the nature of dwarf galaxies. MOND predicts that in the low acceleration regime there should be a high mass deficit  \citep{milgrom19831}. In other words, MOND predicts that if objects are low surface brightness, they should require a lot of dark matter, which is exactly what is found in dwarf galaxies. Tidal dwarf galaxies (TDGs) are also hard to explain in the $\Lambda$CDM paradigm as their formation process does not predict the dragging of large amounts of dark matter away from their host galaxy  (see for example Section 6.5.4 of \citet{famaeyreview} and references therein). The  formation of TDGs is more easily explained in MOND \citep{TDGs}. For a more detailed review of the successes of MOND, we refer the reader to \citet{famaeyreview}.

Despite the success of MOND with regards to galaxy dynamics, MOND is not without its issues. Firstly, to be a viable gravity theory there must be a relativistic formulation, analogous to Einstein's general relativity.  There have been a few proposed relativistic theories that produce MOND as a non-relativistic limit including, for example, Tensor-Vector-Scalar gravity (TeVeS) \citep{teves}, the Dark Fluid model \citep{darkfluid3} and Bimetric MOND \citep{bimond}. However, these have  their limitations and do not rival the success of general relativity.

Another major problem for MOND, and the topic of this paper, is the limited success when trying to describe the mass deficit in clusters of galaxies; see for example \citet{sanders1999} and \citet{sanders2003}. It has been well documented that the inferred mass by MOND in galaxy clusters is not enough to keep the cluster in hydrostatic equilibrium. This result is enough for some to abandon MOND as a viable theory. It is therefore the responsibility of those who study MOND to provide a suitable explanation as to why this missing mass problem arises.  The successful predictions that were made using MOND have led some to believe that the inability to reproduce the galaxy cluster data is just a prediction that there exists some additional mass component that we are yet to detect. One possibility is that MOND and dark matter of some kind are not mutually exclusive and there exist both a breakdown of Newtonian dynamics in the small acceleration regime and some kind of elusive particle, increasing the available mass budget in clusters. This matter could be in the form of 2 eV neutrinos, which was suggested in \citet{sanders2003}. However, analysis of a sample of galaxy clusters by \citet{angus20081} found that the 2 eV neutrino was able to explain the mass deficit in the outer regions of the cluster with missing mass still remaining in the central regions. This would warrant the existence of a different type of neutrino. An 11 eV sterile neutrino was investigated by \citet{angus2009}, which was found to be sufficient to explain the galaxy cluster issue.  Further investigation into solving galaxy clusters in MOND with the aid of neutrinos was conducted in \citet{angusneutrino1} and \citet{angusneutrino2}. Cosmological simulations showed that it was possible to form clusters using hot dark matter (sterile neutrinos). However, the results showed an underabundance of low mass clusters (attributed to the resolution of the simulation) and an overabundance of high mass clusters. They go on to explain how the 11 eV neutrino was ruled out as a neutrino mass of 30-300 eV  was required to produce enough low mass clusters.

Neutrinos  have not been the only attempt to reconcile clusters with MOND. Zhao and Li investigated a family of gravity models controlled by a vector field (see for example \citet{zhaolidarkfluid1} and \citet{zhaolidarkfluid2}. This family of solutions can reproduce different types of behaviour depending on the model parameter choices.  This is based on the assumption that there exists some dark fluid, described by the vector field, which is the source of deviation from Newtonian physics. Another attempt to solve the cluster problem proposed by \citet{darkfluid1} also postulates there exists a dark fluid that permeates space. This idea is motivated by the success of MOND on the galactic scale, but the success of $\Lambda$CDM in cosmological and cluster scales. This dark fluid formalism aims to recover galaxy scale dynamics via a MOND-like phonon mediated force resulting from the dark matter in its superfluid phase. Clusters, in this scenario, would be dominated by the normal phase of the superfluid, which would behave like particle dark matter. This aims for a best of both worlds scenario with regards to dark matter and MOND. There has also been a theory proposed by \citet{dipolarDM}, which includes a substance called dipolar dark matter. This mechanism works on the principle that we do not actually see a breakdown of a gravity law when we empirically see MOND, but what we are witnessing is an enhancement of gravity due to dipole moments of this exotic dark matter aligning with the gravitational field. All these theories have pros and cons, but are not studied in this work. 

Although the aforementioned solutions may hold the keys to the cluster success in MOND, the need for dark matter of any kind has left some unsettled, as the elegance of MOND, in galaxies, is the ability to account for all observations from visible matter only. Therefore aside from these dark fluid-like theories, it has also been noted that if the value of $a_{0}$ were to be increased in clusters, the mass discrepancy could be alleviated. In the same manner that MOND is a more general form of Newtonian dynamics, perhaps MOND in its current form is not universal, but part of a larger gravitational law. The generalization of MOND should not be ruled out hastily given the inherently empirical nature of its formulation. MOND was devised by studying rotation curves, as information of galaxy clusters was limited. Any new adaptation of MOND that increases $a_0$ in clusters would have to somehow shield galaxies, where MOND works well, from the effects of this varying acceleration scale. This idea was formally outlined by \citet{EMOND} in the form of Extended Modified Newtonian Dynamics or EMOND.

Another issue in MOND is the newly discovered ultra diffuse galaxies (UDGs) \citep{ultradiffuse1}, \citep{ultradiffuse2}. Ultra-diffuse galaxies are extremely low density galaxies, residing in the potential wells of galaxy clusters, which require  a lot of dark matter. This required dark matter could be explained in MOND if it were not for the high external field exhibited on the UDGs by the galaxy cluster. The external field of the cluster raises the acceleration in the UDG, making the MOND prediction closer to Newtonian. This could be counteracted if the value of $a_{0}$ were in fact larger in these environments as deviations from Newtonian gravity could occur at higher accelerations. The fact that a higher value of $a_{0}$ in galaxy clusters could simultaneously explain the mass deficit problem and the observations of these UDGs is interesting and should thus be explored in more detail.

This paper is outlined as follows. Section \ref{MONDSection} outlines the MOND formalism and briefly reviews the missing mass problem in clusters with MOND. This is followed by a review of the EMOND formalism. Section \ref{SampleSection} discuses the galaxy cluster sample that we have used throughout the paper. Section \ref{EMONDSection} discusses how well EMOND works in explaining the missing mass problem. We find that EMOND has partial success in tackling this problem and thus in Section \ref{AMONDSection} we try to find an empirical relation, based on the MOND paradigm, which can account for the cluster mass deficit. We find a potential relation, with some success as well, which we compare to the Navarro-Frenk-White
(hereafer NFW) fits described in \citep{sample} in Section \ref{AMONDNFWSection}. In Section \ref{AMONDLimitationSection} we discuss the limitations of this empirical relation and we conclude in Section \ref{Conclusion}.  

\section{Summarizing the MOND-cluster problem }\label{MONDSection}

\subsection{The MOND paradigm and its limitations}

We begin by briefly reviewing the MOND paradigm and its governing equations.  The general motivation for MOND the recognition that the Newtonian prediction of gravity only seemed to fail in low acceleration environments. If there is no non-baryonic dark matter to explain this discrepancy, the gravitational law governing the Universe would have to deviate from Newtonian in these low acceleration environments.

In Newtonian physics, the gravitational field $\Phi$ is determined from the matter distribution $\rho$ via Poisson's equation,

\begin{equation}
\nabla^{2} \Phi = 4 \pi G \rho.
\end{equation}

Determining the gravitational field predicted by MOND requires an alternate Poisson equation \citep{bekenstein1984},

\begin{equation}\label{MONDPoiss}
\nabla \cdot \left[ \mu\left( \frac{|\nabla \Phi|}{a_{0}}  \right) \nabla\Phi\right] = 4 \pi G \rho,
\end{equation}

\noindent
where $a_{0}$ is a universal scale acceleration and $\mu(x)$ is the MOND interpolation function that behaves such that,

\begin{equation}\label{intlimits}
\mu(x) = \begin{cases}

  1 & \text{for  $ x \gg 1$} \\

  x & \text{for $x \ll 1$}

\end{cases}.
\end{equation}

The first case of Equation \ref{intlimits} is the Newtonian regime and the second is known as the deep-MOND regime. The Newtonian Poisson equation is recovered in the $x>>1$ case of Equation \ref{intlimits}. One should note here that the centres of galaxy clusters have strong gravity, meaning they have an $x>1$ MOND value. 

 There are many possible functional forms for $\mu(x)$ that satisfy the criteria of Equation \ref{intlimits}. We opted to choose the simple interpolation function \citep{simplemu},

\begin{equation}\label{simple}
\mu(x) = \frac{x}{1 + x}
,\end{equation}

\noindent
for our analysis.

\begin{figure}
\includegraphics[scale=0.5]{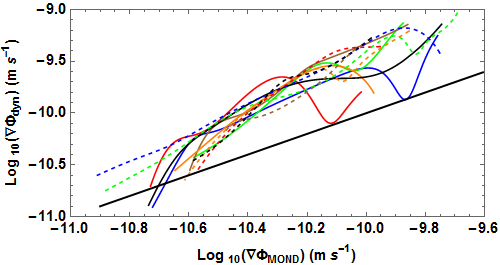}
\caption{Plot showing how the dynamical (or total) gravitational acceleration of the cluster compares to the gravitational acceleration predicted by MOND with a 1:1 line overplotted for illustrative purposes. Each line is a different cluster. We use 12 clusters from \cite{sample}.  As the clusters lie above the 1:1 line (black solid line on plot), the required gravity is more than the `MOND predicted' gravity.}
\label{FF}
\end{figure}

The MOND paradigm has always struggled with rectifying the mass discrepancy in galaxy clusters without including an extra mass component such as neutrinos or non-luminous baryonic matter.  In its usual form MOND is unable to boost the gravitational acceleration in clusters sufficiently whilst this framework is still able to match observational constraints provided by galaxy dynamics. The reason for MOND's inability to accomplish this is that the gravitational acceleration in galaxy clusters is relatively high, in general x>1, so the MOND boost to gravity is weak. However, there is an observed mass deficit in galaxy clusters that would require the cluster to be in the deep-MOND regime ($x<<1$) to rectify with MOND alone.  We demonstrate this issue briefly by applying MOND to the cluster sample described in Section \ref{SampleSection}. We can plot the dynamical acceleration (See Section \ref{SampleSection} for details on how the dynamical acceleration is calculated) for each cluster against the corresponding MOND computed acceleration from Equation \ref{MONDPoiss}; see (Figure \ref{FF}).  If MOND were a universal law that applied to galaxy clusters, Fig \ref{FF} should show 1:1 correlation (within observational errors), by which we mean that the gravitational acceleration predicted by the MOND formula should match the gravitational acceleration predicted by the dynamics of the X-ray gas.  In Figure \ref{FF}, we see that the required gravitational acceleration, derived from the dynamics of the X-ray gas, exceeds the gravitational acceleration predicted by the MOND formula. Thus MOND is under-predicting the gravitational acceleration of each cluster. As mentioned previously, this offset is often reconciled in the MOND paradigm by the presence of some form of non-luminous baryonic matter or neutrinos, which has yet to be detected.

\subsection{Can MOND be generalized to account for clusters?}

\begin{figure}
\includegraphics[scale=0.6]{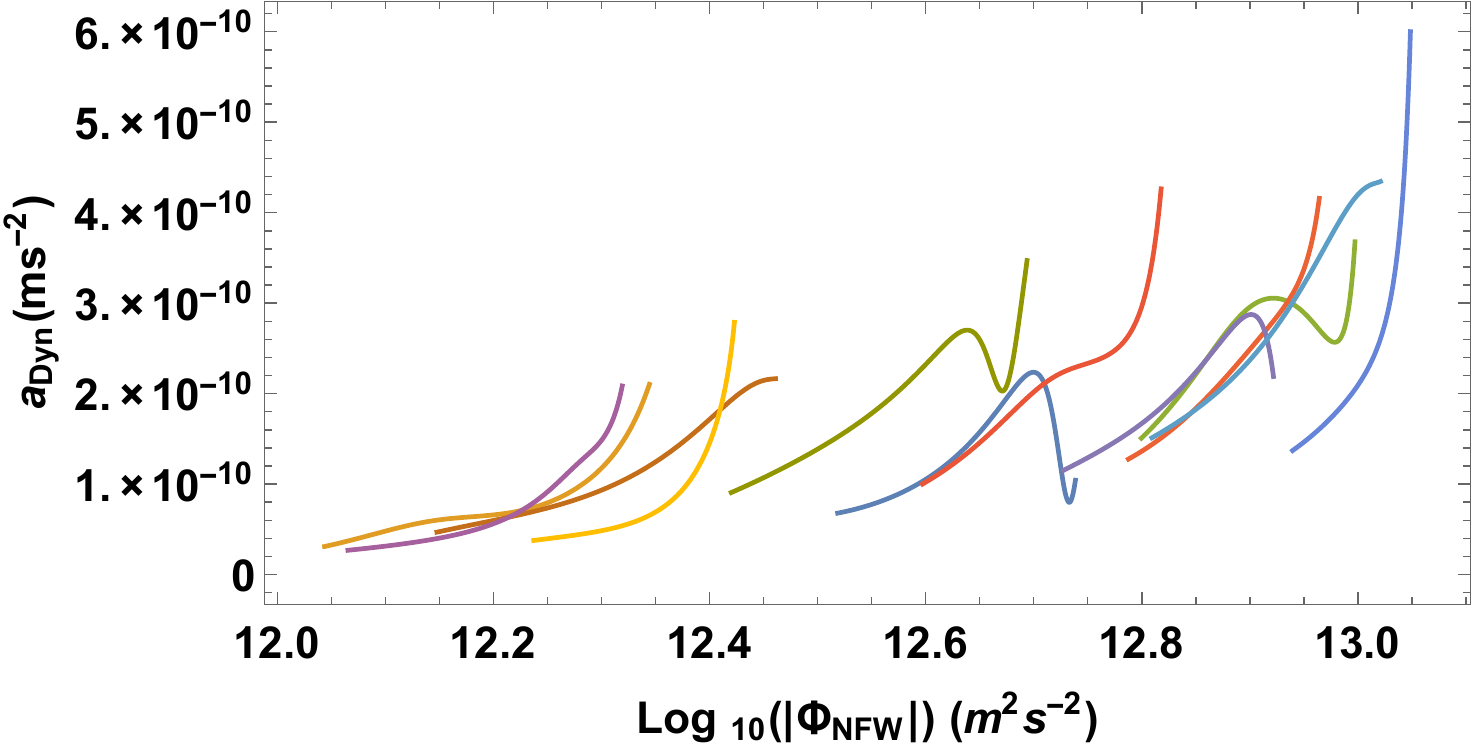}
\caption{Plot showing how the dynamical gravitational acceleration varies with NFW potential for each cluster in our sample. NFW profiles were taken from \cite{sample}. Note the regular trend between the clusters, which suggests a modified gravity law dependent on gravitational potential.}
\label{FPhi}
\end{figure}

As the simple MOND formulation does not hold true for galaxy clusters, efforts were made to try and generalize MOND. In order to do this, one must identify what, if any, are the critical differences between galaxies and galaxy clusters. One such difference is that the gravitational potential is much larger in galaxy clusters than in galaxies. In its current form, MOND has a functional dependence solely on gravitational acceleration. Perhaps a more universal gravity law requires gravitational potential as a mediator of total gravity. Again, using the cluster sample described in Section \ref{SampleSection}, we can see how the gravitational acceleration of the cluster varies as a function of gravitational potential (see Figure \ref{FPhi}).  In Figure \ref{FPhi}, we made the approximation that the gravitational potential of the cluster is equal to the contribution of the dark matter, described by a NFW profile \citep{NFWpaper}, \citep{zhaoprofile}. The NFW profile for each cluster was given in \citet{sample}. We made this approximation as the dark matter NFW profile is analytical and well defined,  whereas computing the gravitational potential from the dynamical data requires us to know  where the edge of the cluster lies in order to truncate the mass. Furthermore, we would have to perform an unnecessary additional step of numerically integrating the dynamical acceleration to find the potential. This integration would involve a boundary value that is dependent on the truncation radius of the cluster. As we are only interested in getting an intuition of how the properties of the galaxy clusters correlate with gravitational potential, the NFW approximation seems to be reasonable compared to performing the integration of the dynamical acceleration.

This trend with potential was the main motivation of the Zhao and Famaey Extended MOND (EMOND) formalism \citep{EMOND}. In EMOND, the corresponding Poisson equation is written as

\begin{equation}\label{EMONDPoiss}
4 \pi G \rho = \nabla \cdot \left[\mu\left( \frac{|\nabla \Phi|}{A_{0}(\Phi)}  \right) \nabla \Phi  \right] - T_{2},
\end{equation}

\noindent where

\begin{equation}
T_{2} = \frac{1}{8\pi G}\left| \frac{d (A_{0}(\Phi))^{2}}{d\Phi} \right|\left[ y F'(y) - F(y) \right]
,\end{equation}
 where $A_{0}(\Phi)$ is a functional alternative to the MOND constant $a_{0}$  with units of acceleration, and $\mu$ is the same MOND interpolation function, $dF(y)/dy = \mu(\sqrt{y})$  and $y = |\nabla \Phi|^{2}/A_{0}(\Phi)^{2}$. It is immediately apparent that Equation \ref{EMONDPoiss} is much more complicated than the regular MOND Poisson equation owing to the second term. It has however been argued that in certain situations, such as galaxy clusters, this term is small compared to the first \citep{EMOND}, giving an approximate expression,

\begin{equation}\label{EMONDPoiss2}
4 \pi G \rho \approx \nabla \cdot \left[\mu\left( \frac{|\nabla \Phi|}{A_{0}(\Phi)}  \right) \nabla \Phi  \right]
,\end{equation} 

\noindent
for the EMOND gravitational potential. In Section \ref{EMONDSection} we explicitly show that we are just in making this approximation. Equation \ref{EMONDPoiss2}, the approximate EMOND Poisson equation, is equivalent to Equation \ref{MONDPoiss} with a modified scale acceleration. The idea behind Equation \ref{EMONDPoiss2} is to increase the MOND acceleration scale in high potential environments. This decreases the value of the MOND interpolation function, $\mu$, thus increasing the total gravitational acceleration that is predicted by MOND ($a_{MOND} \approx a_{Newt}/\mu$). In order to solve the EMOND Poisson equation, we must define the behaviour of $A_{0}(\Phi)$, which we discuss in Section \ref{EMONDSection}. Although we determine the exact behaviour of $A_{0}$ in Section \ref{EMONDSection}, we should provide some understanding of the general desired behaviour that we hope to emulate. From our understanding of galaxy rotation curves in MOND, we know that in general galaxies are well described by a constant $a_{0} \approx 1.2 \times 10^{-10}$ ms$^{-2}$. We therefore want to preserve this behaviour. Therefore the standard $a_{0}$ should be a limit of $A_{0}$ in the low potential regime such that $A_{0}(\Phi) \rightarrow a_{0}$ as $\Phi \rightarrow 0$.  

\subsection{The external potential effect}\label{ExtPot}

As the modification to the MOND formula we discuss here relies on the gravitational potential, even a constant gravitational potential across a system from an external source affects the dynamics. This feature can be thought of as an external potential effect \citep{EMOND}, which is analogous to the external field effect (EFE) that arises in regular MOND; see for example \citet{EFE1} for work on the EFE in globular clusters; \citet{EFE2}, which looks at the EFE in the solar system; and \citet{decliningcurves}, for work with the EFE and declining rotation curves in MOND. The treatment of the external potential effect should be fully examined in future work. The original EMOND work assumes that the main contribution to the external potential of the Milky Way can be attributed to the Virgo cluster. Work on the escape velocity of the Milky Way \citep{escapingfrommond} calculates the external potential for the Milky Way to be approximately $\Phi_{ext} \approx -10^{-6} c^{2}$. However, the typical gravitational potentials of our galaxy clusters are much larger than this so we are able to exclude this effect from our calculations. The importance of the external potential effect is most prominent in galaxies and objects that lie within clusters or near objects with large gravitational potentials.  For example, we mentioned earlier that there has recently been a discovery of many so-called ultra diffuse galaxies (UDGs) inside galaxy clusters. The existence of these objects could pose some threat to MOND as the external field from the clusters should make the dynamics inside UDGs closer to Newtonian ($\mu(x)\rightarrow 1$). However, if this external potential effect was present, the fact that the UDGs are within the deep potential well of the galaxy clusters could cause UDGs to show MOND-like behaviour or at least exhibit stronger gravity than that expected from Newtonian physics. The external potential effect could be a natural explanation as to why objects exist within clusters that behave similar to UDGs. However, recent estimates of dark matter haloes in these objects show an extremely high dark to baryonic mass ratio \citep{dragonfly44}. One would need to check that the external potential effect required to reconcile that the UDGs does not conflict with the host galaxy cluster.

\section{Analysing a galaxy cluster sample}\label{SampleSection}

To begin, we first need to select a galaxy cluster sample. We use a Chandra galaxy cluster sample \citep{sample} that provides analytical fits to the temperature and density profiles of the X-ray gas. We use 12 clusters from the sample. The clusters are nearby (z<1) and are in the mass range $\approx 10^{14}-10^{15}$M$_{\odot}$. For a thorough analysis, we also need to model the brightest cluster galaxy (BCG) component. The BCG modelling is outlined in Section \ref{BCGsec}. 

\subsection{Cluster model}

To model the X-ray gas density \citet{sample} chooses an analytical expression to which the data is fit. This model is composed of three parts. The first part models the cuspy core usually found in relaxed clusters and is achieved by a power law density profile. They then include an additional term that models the steepening brightness profile at $r > 0.3 r_{200}$, where $r_{200}$ is the radius at which the average density of the cluster falls to 200 times that of the critical density of the Universe. They finally add another power law component to allow the density profile to be very general for fitting freedom. Combining these three components leads to their overall emission profile for the X-ray gas,

\begin{equation}\label{density}
n_{e}n_{p}(r) = n_{0}^{2} \frac{\left(r/r_{c}\right)^{-\alpha}}{\left(1 + \left( r/r_{c} \right)^{2}\right)^{3\beta - \alpha/2}}\frac{1}{\left(1 + \left( r/r_{s} \right)^{\gamma}\right)^{\epsilon/\gamma}} +  \frac{n_{0}^{2}}{\left(1 + \left( r/r_{c2} \right)^{2}\right)^{3\beta_{2}}}.
\end{equation}

\noindent
where $r_{c}$, $r_{c2}$, and $r_{s}$ are all scale radii and $\alpha$, $\beta$, $\gamma$, $\epsilon$, and $\beta_{2}$ are all dimensionless parameters. $n_{e}$ and $n_{p}$ are electron and proton number densities, and $n_{0}$ is a parameter that takes the meaning of a central number density.  The emission profile is converted into the mass density by $ \rho_{g}(r) \approx 1.252 m_{p} \left(n_{p} n_{e} \right)^{1/2}$,  where $m_{p}$ is the mass of a proton. The parameter fits for each galaxy cluster are given in Table \ref{densityparams}, which were taken from \citet{sample}.

\begin{table*}
\centering
\caption{Table of parameters as given in \citet{sample} for the X-ray gas emission profile described by Equation \ref{density}. We omitted one cluster from our analysis as \citet{sample} do not provide NFW fits for dark matter, so we cannot compare our result to that in a consistent manner. $\gamma$ for each cluster was chosen to be 3.0. Values in table labelled N/A refer to the clusters for which \citep{sample} did not provide parameters, making those clusters a two-component density model rather than three (see Equation \ref{density}). }
\label{densityparams}
\begin{tabular}{|l|l|l|l|l|l|l|l|l|l|}
\hline
Cluster      & $n_{0}$   & $r_{c}$    & $r_s$    & $\alpha$ & $\beta$  & $\epsilon$ & $n_{02}$    & $r_{c2}$    & $\beta_{2}$ \\
&($10^{-3}\text{cm}^{-3}$)&(kpc) &(kpc)  & & & & ($10^{-1}\text{cm}^{-3}$) &(kpc)&\\
\hline
A133         & 4.705  & 94.6 & 1239.9 & 0.916 & 0.526 & 4.943   & 0.247 & 75.83 & 3.607    \\
A262         & 2.278  & 70.7  & 365.6  & 1.712 & 0.345 & 1.760   & N/A   & N/A   & N/A     \\
A478         & 10.170  & 155.5 & 2928.9 & 1.254 & 0.704 & 5.000 & 0.762 & 23.84    & 1.00       \\
A1413        & 5.239 & 195.0 & 2153.7 & 1.247 & 0.661 & 5.000   & N/A   & N/A   & N/A     \\
A1795        & 31.175 & 38.2  & 682.5 & 0.195  & 0.491 & 2.606   & 5.695   & 3.00   & 1.00     \\
A1991        & 6.405  & 59.9  & 1064.7  & 1.735 & 0.515 & 5.000  & 0.007 & 5.00     & 0.517   \\
A2029        & 15.721 & 84.2  & 908.9 & 1.164 & 0.545 & 1.669    & 3.510 & 5.00     & 1.00       \\
RXJ1159+5531 & 0.191  & 591.9 & 640.7  & 1.891 & 0.838 & 4.869   & 0.457 & 11.99 & 1 .00      \\
MKW4         & 0.196   & 578.5 & 595.1 & 1.895 & 1.119 & 1.602   & 0.108 & 30.11 & 1.971  \\
A383             & 7.226 & 112.1 & 408.7 & 2.013 & 0.577 & 0.767 &     0.002 & 11.54 & 1.00  \\
A907             & 6.252 & 136.9 & 1887.1 & 1.556 & 0.594 & 4.998 &    N/A  & N/A    & N/A   \\
A2390            & 3.605 & 308.2 & 1200.0 & 1.891 & 0.658 & 0.563 &    N/A & N/A      & N/A  \\
\hline 
\end{tabular}
\end{table*}

We also require the temperature profile of the gas as we are choosing to relax the common assumption that the galaxy clusters are isothermal. Again, we follow the profile provided by \citet{sample},

\begin{equation}\label{tempprofile}
T(r) = T_{0} \hspace{2mm} \frac{\left( r/r_{cool} + T_{min}/T_{0}\right)}{(r/r_{cool})^{a_{cool}}+1}\hspace{2mm} \frac{\left( r/r_{t}\right)^{-a}}{\left((r/r_{t})^{b}+1\right)^{c/b}},
\end{equation}

\noindent
which accounts for the cooling of the X-ray gas in the central regions of the cluster, which according to \citet{sample} may be the result of radiative cooling. The parameter $T_{min}$ is the central temperature, $T_{0}$ is a scale temperature, $r_{cool}$ and $r_{t}$ are scale radii, and $a_{cool}$, a b and c are dimensionless parameters. Outside the central, cooler region of the cluster the temperature is described by a broken power law.  The parameter fits for the temperature can be found in Table \ref{tempparams}, again taken from \citet{sample}.

\begin{table*}
\centering
\caption{Table of gas temperature parameters as given in \citep{sample} for Equation \ref{tempprofile}. }
\label{tempparams}
\begin{tabular}{|l|l|l|l|l|l|l|l|l|}
\hline
Cluster      & $T_{0}$    & $r_{t}$   & a     & b    & c   & $T_{min}/T_{0}$ & $r_{cool}$ & $a_{cool}$ \\
&kev&Mpc&&&&&kpc&\\
\hline
A133         & 3.61  & 1.42 & 0.12  & 5    & 10  & 0.27    & 57    & 3.88  \\
A262         & 2.42  & 0.35 & -0.02 & 5    & 1.1 & 0.64    & 19    & 5.25  \\
A478         & 11.06 & 0.27 & 0.02  & 5    & 0.4 & 0.38    & 129   & 1.6   \\
A1413        & 7.58  & 1.84 & 0.08  & 4.68 & 10  & 0.23    & 30    & 0.75  \\
A1795        & 9.68  & 0.55 & 0     & 1.63 & 0.9 & 0.1     & 77    & 1.03  \\
A1991        & 2.83  & 0.86 & 0.04  & 2.87 & 4.7 & 0.48    & 42    & 2.12  \\
A2029        & 16.19 & 3.04 & -0.03 & 1.57 & 5.9 & 0.1     & 93    & 0.48  \\
RXJ1159+5531 & 3.74  & 0.1  & 0.09  & 0.77 & 0.4 & 0.13    & 22    & 1.68  \\
MKW4         & 2.26  & 0.1  & -0.07 & 5.00 & 0.5 & 0.85    & 16    & 9.62 \\
A383             & 8.78  & 3.03 & -0.14 & 1.44 & 8.0 & 0.75    & 81    & 6.17 \\
A907             & 10.19 & 0.24 & 0.16  & 5.00 & 0.4 & 0.32    & 208   & 1.48 \\
A2390            & 19.34 & 2.46 & -0.10 & 5.00 & 10.0& 0.12    & 214   & 0.08 \\
\hline 
\end{tabular}
\end{table*}

\subsubsection{BCG model}\label{BCGsec}

We also model the brightest cluster galaxy (BCG) for each cluster, which affects the central parts of the cluster. This is an important aspect of the model as it is in the central regions of the clusters where the mass deficit is greatest. To model the BCG, it is reasonable to make the approximation that it is spherical, with a density following the well-known Hernquist profile, 

\begin{equation}
\rho_{BCG} (r) = \frac{M h}{2\pi r \left(  r + h \right)^{3}}
,\end{equation}

\noindent where M is the BCG mass and h is the BCG scale length.  We assign a BCG mass proportional to the overall mass of the cluster such that $M_{BCG} = 5.3 \times 10^{11} \left( \frac{M_{500}}{3\times 10^{14} M_{\odot}} \right)^{0.42}M_{\odot}$, which follows the work of \citet{BCGmass}. Here, $M_{500}$ is the enclosed mass at $r_{500}$, which is the radius at which the average density is 500 times the critical density of the Universe.  We also assign each cluster a Hernquist scale length of $30$ kpc. Although these estimates may be crude, the BCG should only affect the inner parts of the cluster and thus should not affect our overall conclusions too much.

\subsubsection{Radial range of the data}

The radial range of the X-ray data is limited by two factors: an inner boundary or cut-off radius and an upper bound radius where the X-ray brightness is not detected. \citet{sample} choose an inner radial boundary such that they exclude the central temperature bin and an outer radial boundary where the X-ray brightness is no longer detected above 3$\sigma$ or the limit of the Chandra field of view was reached. Therefore the model is fit to a section of the radial extent of the cluster, between these two limits, and then extrapolated to lower and higher radii. In Table \ref{r500}, we show the inner most radii as given in \citet{sample}. For our plots throughout this paper, we choose not to show any data below this radius.

For the outer radius we choose to extrapolate past the maximum radius of each cluster. We extrapolate to the radius $r_{200}$ for each cluster, which can be crudely calculated via $r_{200} \approx 1.5 r_{500}$ \citep{sample}. The $r_{500}$ values from \citet{sample} are also shown in Table \ref{r500}.

\begin{table}
\centering
\caption{$r_{500}$ values for each cluster as given in \citep{sample}; $r_{200}$ values can be calculated approximately via $r_{200} \approx 1.5 r_{500}$.}
\label{r500}
\begin{tabular}{|l|l|l|}
\hline
Cluster      & $r_{500}$ (kpc) & $r_{min}$ (kpc) \\
\hline
A133         & 1007  & 40\\
A262         & 650  &  10 \\
A478         & 1337 & 30\\
A1413        & 1299 & 20\\
A1795        & 1235 & 40\\
A1991        & 732  & 10\\
A2029        & 1362 & 20\\
RXJ1159+5331 & 700  & 10\\
MKW4         & 634  & 5\\
A383             & 944  & 25\\
A907             & 1096 & 40\\
A2390            & 1416 & 80\\
\hline
\end{tabular}
\end{table}
\subsection{Dynamical mass of the clusters}

In the following sections we  show the results of dynamical masses, potentials, and accelerations so we therefore must define these quantities. When we discuss `dynamical' quantities in this work, we refer to the value calculated by assuming the cluster gas is in hydrostatic equilibrium. This is a common assumption in determining cluster masses. The equation for hydrostatic equilibrium is written as
\begin{equation}\label{hydrostatic}
\frac{d\Phi_{dyn}}{dr} = \frac{G M_{dyn}(r)}{r^{2}} = -\frac{1}{\rho_{g}(r)}\frac{d}{dr}\left[ \frac{\rho_{g} kT(r)}{w m_{p}} \right]
,\end{equation}

\noindent
where $\Phi_{dyn}$ is the dynamical potential, $m_{p}$ is the mass of a proton ($\approx 1.67\times10^{-27}$  kg), and $w$ is the mean molecular weight ($\approx 0.609$). The value $\rho_{g}(r)$ is the gas density, $T(r) $ is the gas temperature, and $M_{dyn}(r)$ is the enclosed dynamical mass at radius r. Using some mathematical prowess, Equation \ref{hydrostatic} can be transformed into

\begin{equation}\label{DynamicalMass}
M_{dyn}(r) = -\frac{kT(r)r}{G wm_{p}}\left[ \frac{d \ln \rho_{g}(r)}{d \ln r} + \frac{d \ln T(r)}{d \ln r} \right].
\end{equation}

In $\Lambda$CDM language, $M_{dyn}$ would be the mass of the baryons and dark matter required to keep the cluster in hydrostatic equilibrium, which we refer to as the dynamical mass. From this, we can calculate the dynamical acceleration,

\begin{equation}\label{DynamicalAcceleration}
a_{dyn}(r) = \nabla \Phi_{dyn}(r) = \frac{G M_{dyn}(r)}{r^{2}}
,\end{equation}

\noindent and thus the dynamical potential can be found by integrating Equation \ref{DynamicalAcceleration},

\begin{equation}\label{DynamicalPotential}
\Phi_{dyn}(r) = \int^{r}_{\infty} a_{dyn}(\tilde{r}) d\tilde{r}.
\end{equation}

In Equation \ref{DynamicalPotential} the limits are chosen such that the gravitational potential is negative.  Because of the choice of gas density and gas temperature profile used to determine the dynamical mass, it is very difficult to determine the dynamical potential of each cluster from Equation \ref{DynamicalPotential}. This is because the dynamical gravitational potential would be sensitive to a choice of cut-off radius, i.e. where we truncate the gas density to zero. The reason for this sensitivity is due to the integral from $\infty$ in Equation \ref{DynamicalPotential}. We would have to make some assumption about how the cluster mass behaves outside $r_{200}$. In order to circumvent this issue, we take note of the fact that the NFW dark matter profile provides good fits to the dynamical mass. The NFW profile has a well-defined analytical expression for gravitational potential. Therefore we make the assumption that $\Phi_{dyn} \approx \Phi_{NFW}$ for each cluster,

\begin{equation}
\Phi_{NFW} \approx \Phi_{dyn} = -\frac{4 \pi G r_{s}^{3} \ln\left[ \frac{r + r_{s}}{r_{s}} \right]\rho_{s}}{r}
\end{equation}

\noindent where $r_{s}$ and $\rho_{s}$ are the scale radius and density, respectively. We should stress at this point that the NFW profile is not an exact solution of the MOND equations and we therefore expect some discrepancies in our our plots. We use this assumption as a guide.

\subsection{Newtonian mass of the clusters}

We can find the Newtonian predicted mass from the cluster by integrating the gas density profile along with the BCG density profile,

\begin{equation}
M_{N}(r) = \int^{r}_{0} 4 \pi \tilde{r}^{2}( \rho_{g}(\tilde{r})+\rho_{BCG}(\tilde{r})) d\tilde{r}.
\end{equation}

We assume in this work that the only significant stellar mass is the BCG mass and that other stellar mass is small compared to the gas mass.  The Newtonian acceleration and potential can then by found by analogous expression of Equation \ref{DynamicalAcceleration} and \ref{DynamicalPotential}.

\section{Applying EMOND to the sample of clusters}\label{EMONDSection}

The initial EMOND theory was qualitatively outlined in \citet{EMOND} but has until now undergone no quantification with regards to applying the formalism to a sample of galaxy clusters. To accomplish this, we first take the approximate spherically symmetric version of Equation \ref{EMONDPoiss2},

\begin{equation}\label{EMONDSpherical}
\nabla \Phi_{N} \approx \mu\left( \frac{|\nabla \Phi|}{A_{0}(\Phi)} \right)\nabla \Phi.
\end{equation}

\begin{figure}
\includegraphics[scale=0.6]{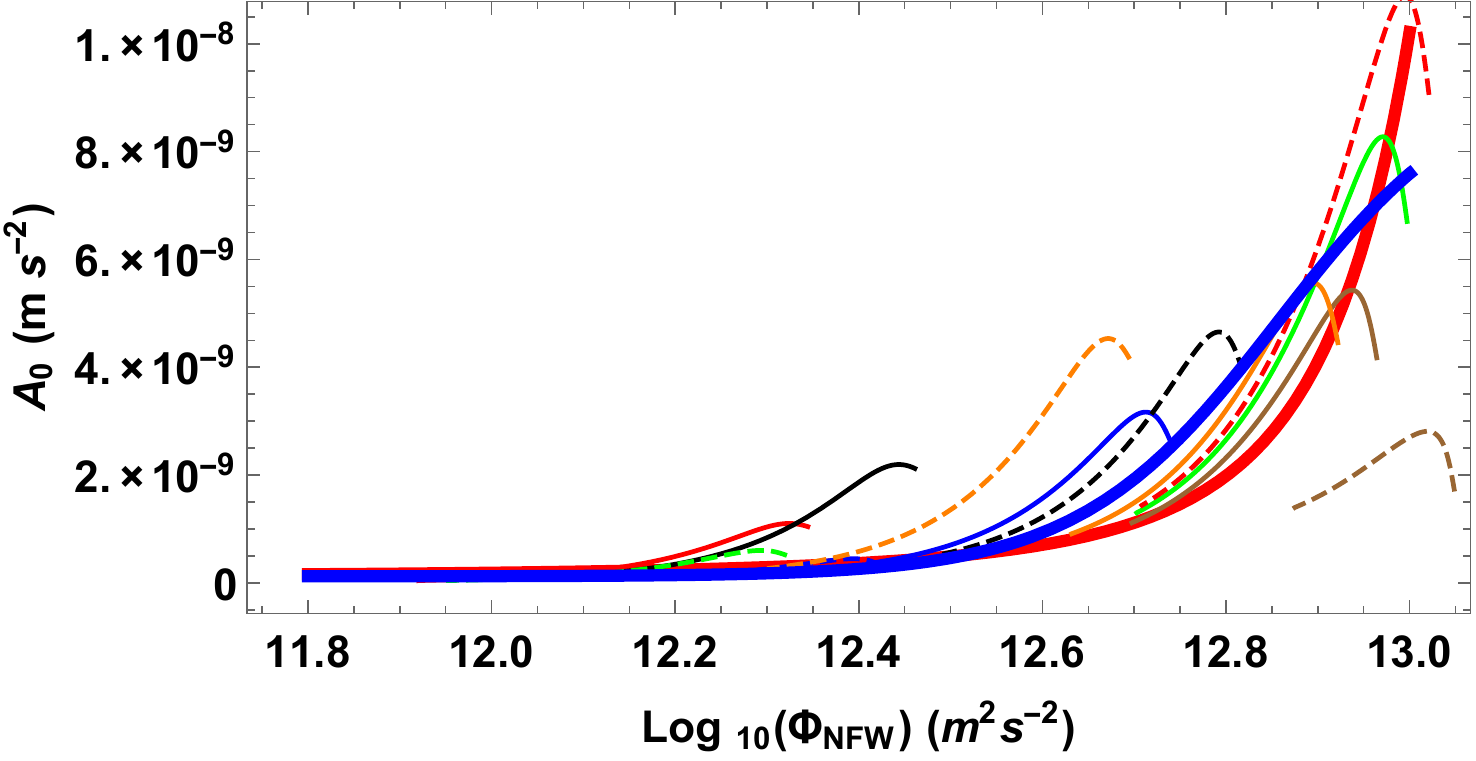}
\caption{Plot showing the required value of $A_{0}$ according to the EMOND formalism (Equation \ref{A0find}) such that predicted gravitational acceleration from EMOND matches the dynamically calculated acceleration (Equation \ref{DynamicalMass}) for each cluster in the sample (thin lines). We plotted this required $A_{0}$ against the NFW gravitational potential as a estimation of the behaviour within the cluster. The shape takes the form of an exponential function overplotted in red and described by Equation \ref{A0exp}. Also plotted in blue is the more complicated $A_{0}$ function described by Equation \ref{A0tan} with a parameter choice, $\Phi_{0}= -2700000^{2}$ m$^{2}$s$^{-2}$ and $A_{0max}= 80a_{0}$.  Note the apparent turnover of $A_0$ at the deepest potential is an artefact because the total gravitational potential is not just that of an NFW profile as assumed here. }
\label{A0plot}
\end{figure}

\noindent We can then invert Equation \ref{EMONDSpherical} to find an expression for $A_{0}(\Phi)$,

\begin{equation}\label{A0find}
A_{0}(\Phi) = \frac{-|\nabla\Phi_{N}| |\nabla\Phi| + |\nabla\Phi|^{2}}{|\nabla\Phi_{N}|},
\end{equation}

\noindent
assuming $\mu$ takes the form of Equation \ref{simple}.  Therefore, we can empirically find out if the data favours the EMOND formalism or whether EMOND cannot explain the missing mass in clusters by determining if there is a single function of $A_{0}(\Phi)$ that can explain the mass discrepancy in all the clusters in our sample.  In Figure \ref{A0plot} we plot the required $A_{0}$, given by Equation \ref{A0find}, as a function of the NFW gravitational potential. To accomplish this, we assume that the total gravitational acceleration is approximated by the dark matter NFW profile for each cluster. We find that there is a general trend of the required $A_{0}$ increasing with gravitational potential, which seems to follow an exponential curve (Figure \ref{A0plot}), for example, 

\begin{equation}\label{A0exp}
A_{0}(\Phi) = a_{0} \exp\left( \frac{\Phi}{\Phi_{0}}\right)
,\end{equation} 

\noindent where $\Phi_{0}$ is the scale potential that we have empirically chosen to be $|\Phi_{0}| \approx 1500000^{2}\text{m}^{2}\text{s}^{-2}$.  One can see that in low potential environments, such as galaxies, that $A_{0}\approx a_{0}$ and thus MOND dynamics are preserved. We plotted Equation \ref{A0exp} in Figure \ref{A0plot} (red line). This functional form has the issue that it rises very quickly in high potential regions. This could cause issues in higher mass clusters. Therefore we tested a second function. 

\begin{equation}\label{A0tan}
A_{0}(\Phi) = a_{0} + (A_{0max}-a_{0})\left[ \frac{1}{2}\tanh \left[ \log \left(\frac{\Phi}{\Phi_{0}}  \right)^{2} \right] +\frac{1}{2}\right],
\end{equation} 

\noindent where the acceleration scale, $A_{0}(\Phi)$, grows like a step function with minimum value $a_{0}$ in low potential regions and maximum value $A_{0max}$ in high potential regions. {\bf For this approach, we are inspired by an alternate modification to gravity that uses a density dependent modification to gravity to try and explain galaxies and galaxy clusters without the need for dark matter \citep{refractedgravity}. In this formulation, gravitational dynamics are described by a modified Poisson equation,
$
\nabla \cdot \left( \epsilon(\rho) \nabla \Phi \right) = 4\pi G \rho,
$
where $\epsilon$ is a dimensionless function of density. In their work, the functional choice of $\epsilon(\rho)$ is a smooth step function such that,
$
\epsilon(\rho) = \epsilon_{0} + (1-\epsilon_{0})\frac{1}{2}\left[ \tanh \left[ \log \left(\frac{\rho}{\rho_{c}}  \right)^{q} \right] +1\right],
$
where $\epsilon_{0}$ and $q$ are free parameters and $\rho_{c}$ is a density scale, which is analogous to our potential scale $\Phi_{0}$.} We overplot this function in Figure \ref{A0plot} (blue line) along with the simple exponential.

\begin{figure}
\includegraphics[scale=0.6]{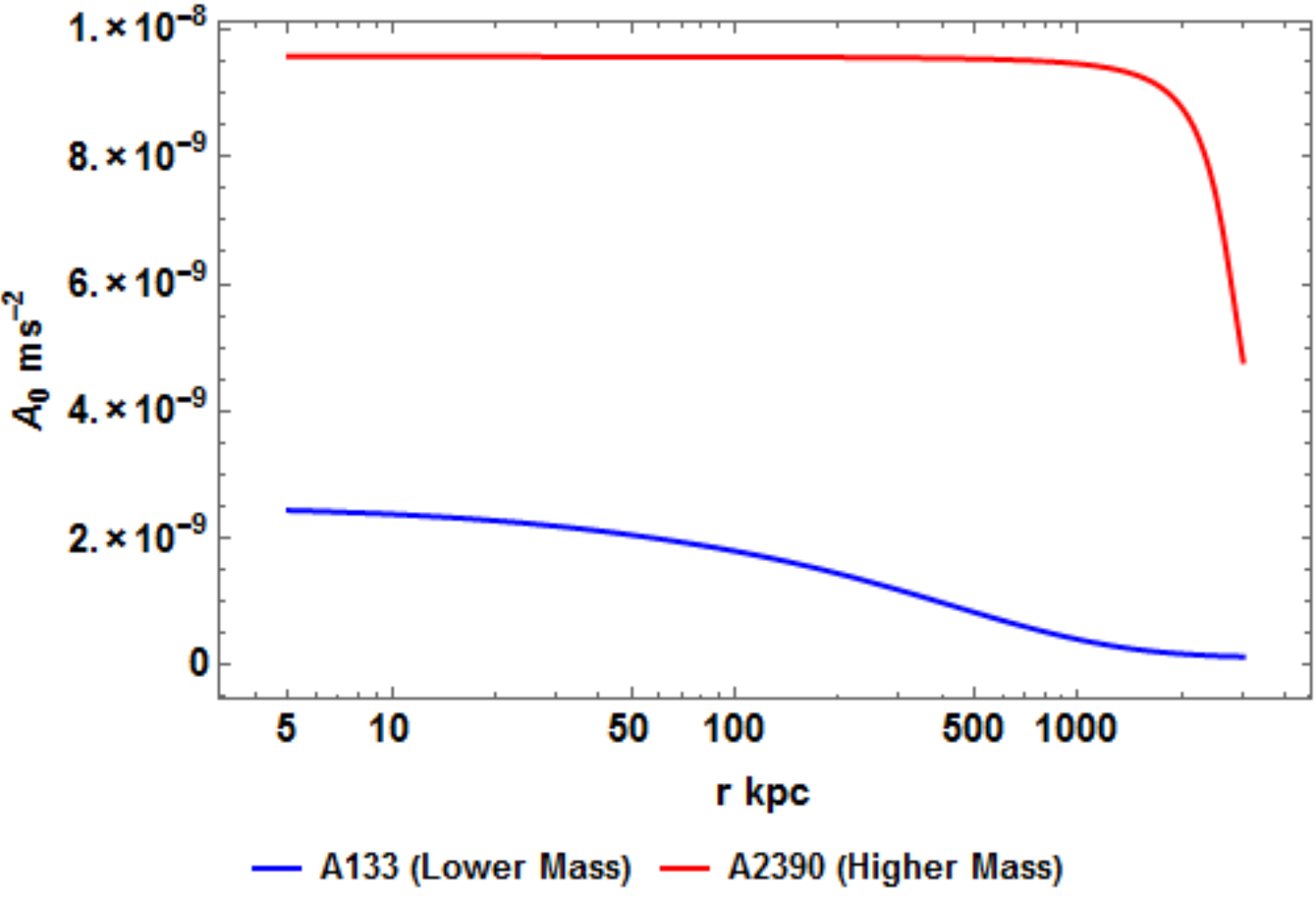}
\caption{Plot showing the calculated $A_{0}$ for 2 clusters, A133 and A2390, using Equation \ref{A0tan}. Note that unlike A133, the massive A2390 has a potential mostly above the step function, hence shows little dependence on the changing of the boundary potential (cf. Figure \ref{MassPlot6}).}
\label{A0clusterplot}
\end{figure}

In Figure \ref{A0clusterplot} we plot the calculated $A_{0}$ as a function of radius for two clusters, A133 and A2390. We choose these two as A133 is a less massive cluster and A2390 is the most massive cluster. We can see that in A2390 the value of $A_{0}$ is mostly at the maximum value that Equation \ref{A0tan} allows as it has a large gravitational potential, whereas A133 has a much lower $A_{0}$ which increases towards the centre.

\begin{figure*}
\begin{tabular}{ccc}
\includegraphics[scale=0.5]{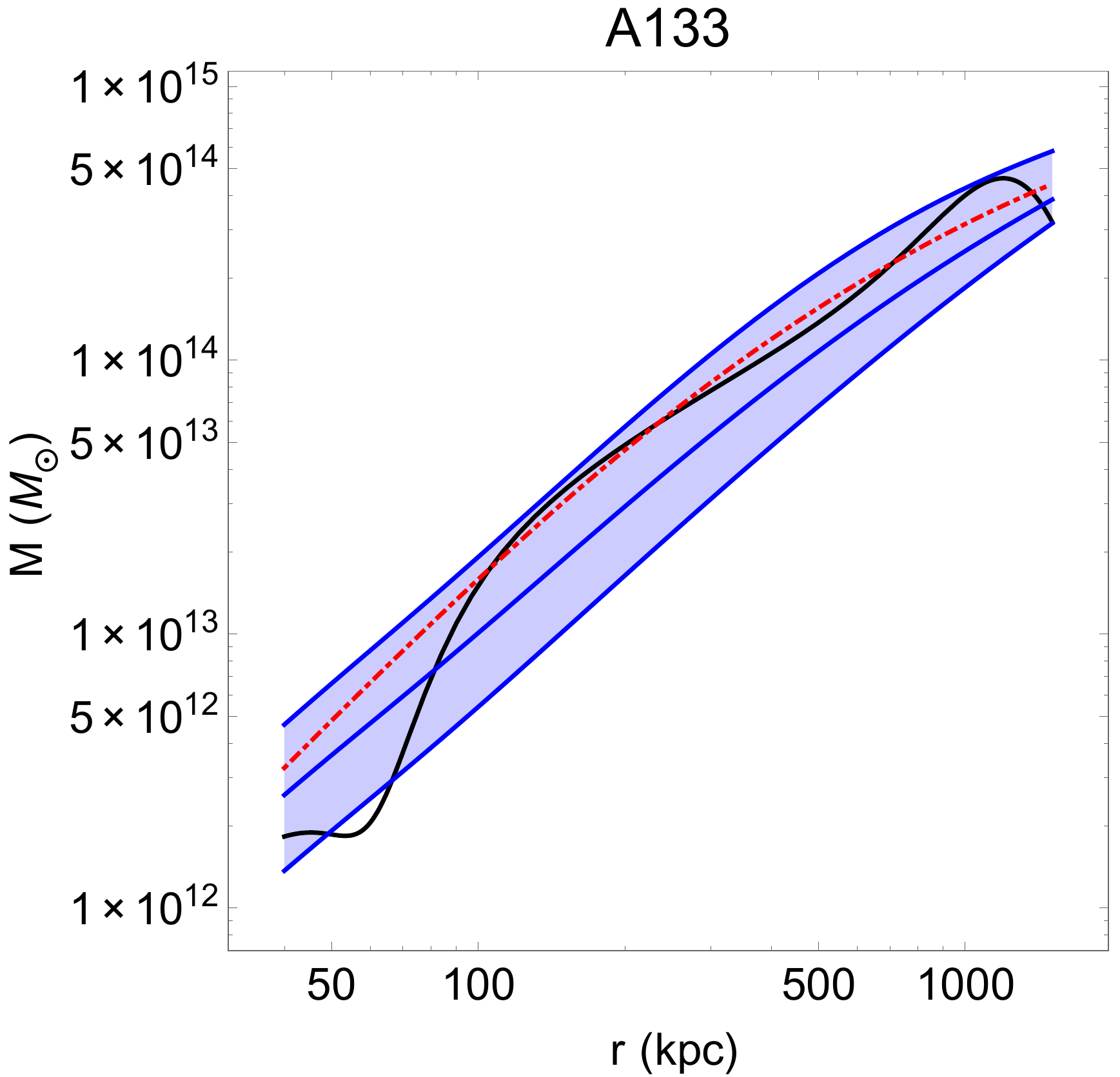} & \includegraphics[scale=0.5]{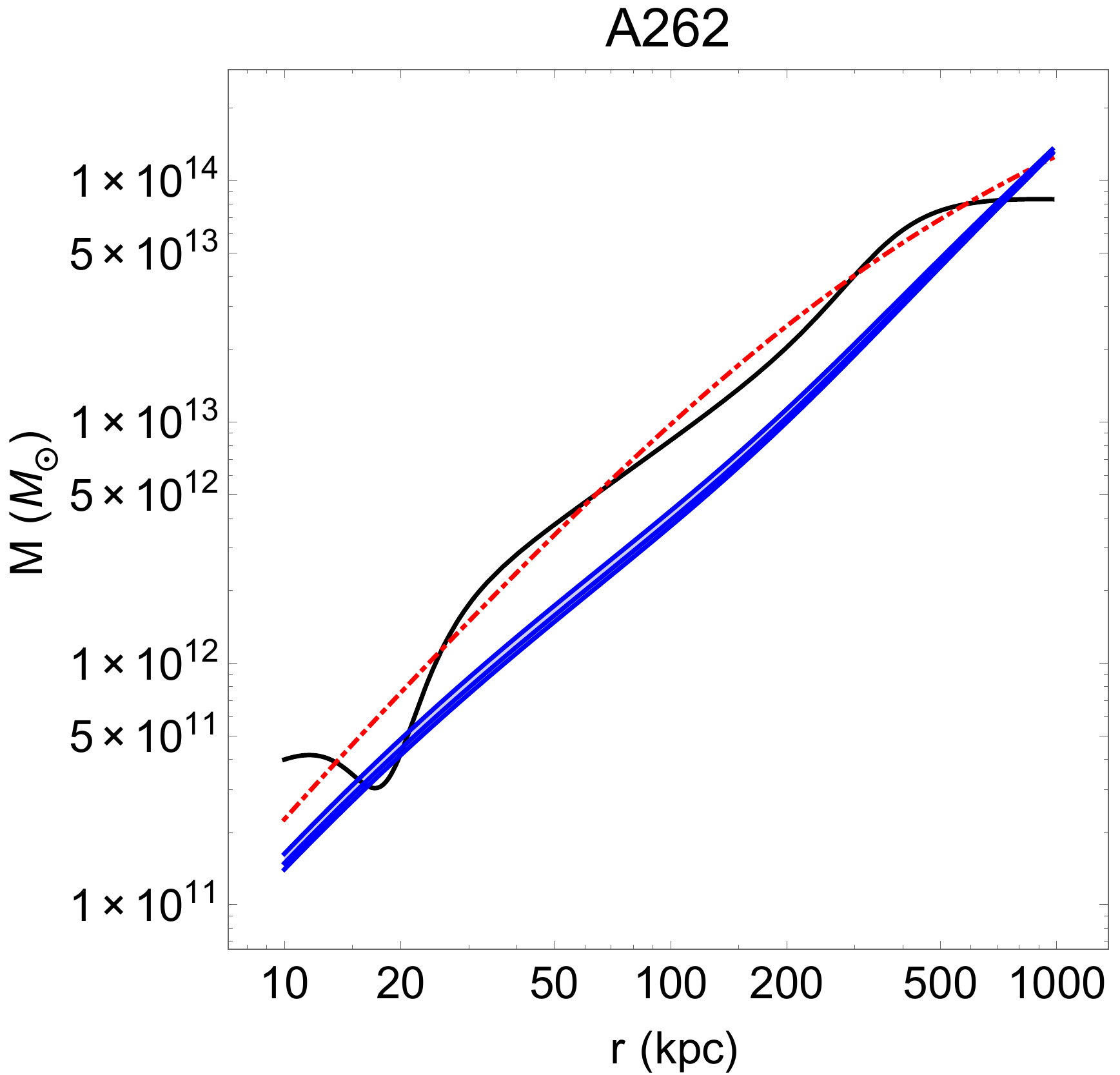}
\end{tabular}
\caption{Plot showing mass profiles for the EMOND relation for different values for the boundary potential (blue shaded region). Also the plot shows the NFW (red dotted line) and dynamical masses (black solid line) for clusters A133 and A262.}
\label{MassPlotEMOND}
\end{figure*}

\begin{figure*}
\begin{tabular}{ccc}
\includegraphics[scale=0.5]{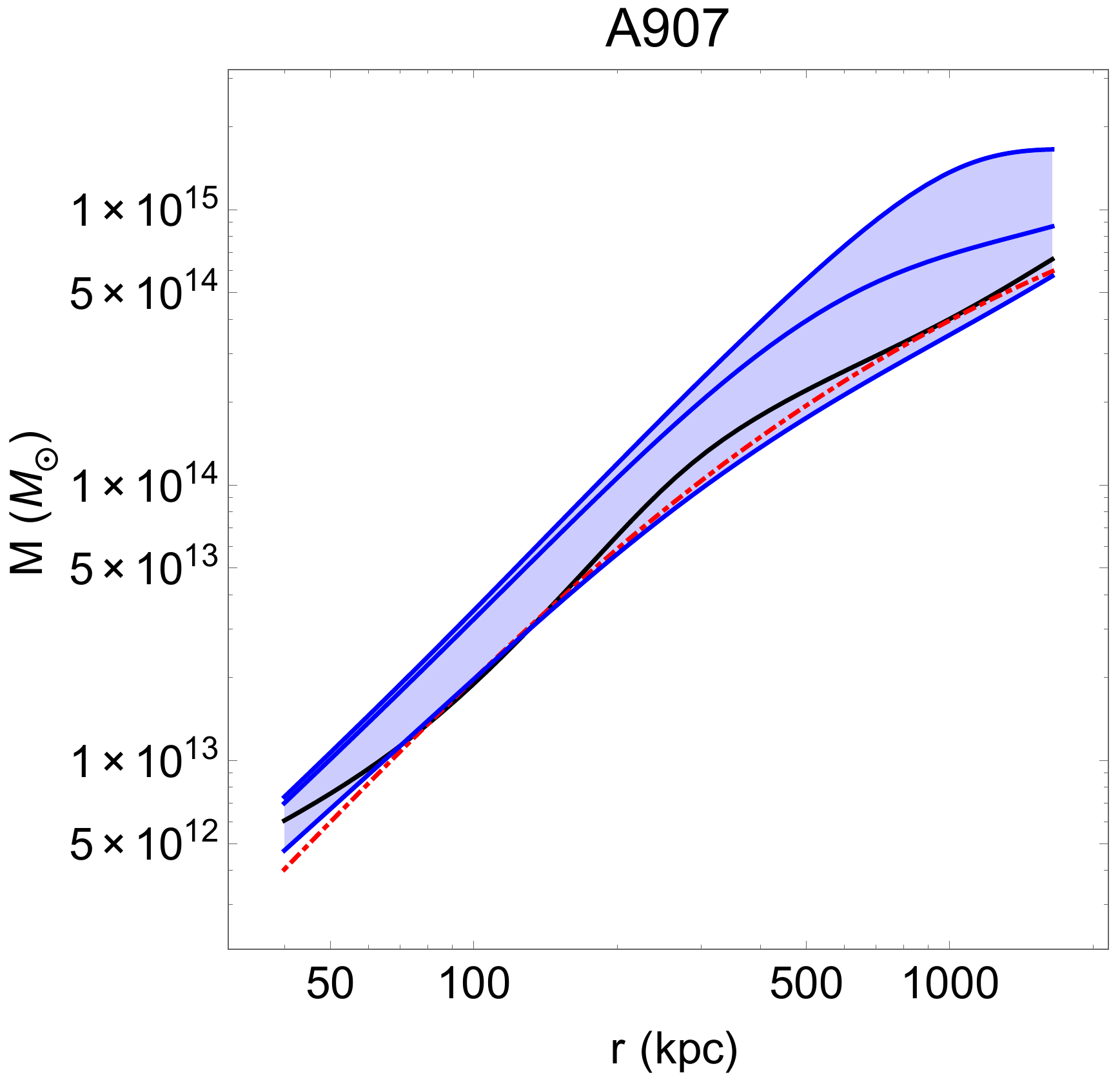} & \includegraphics[scale=0.5]{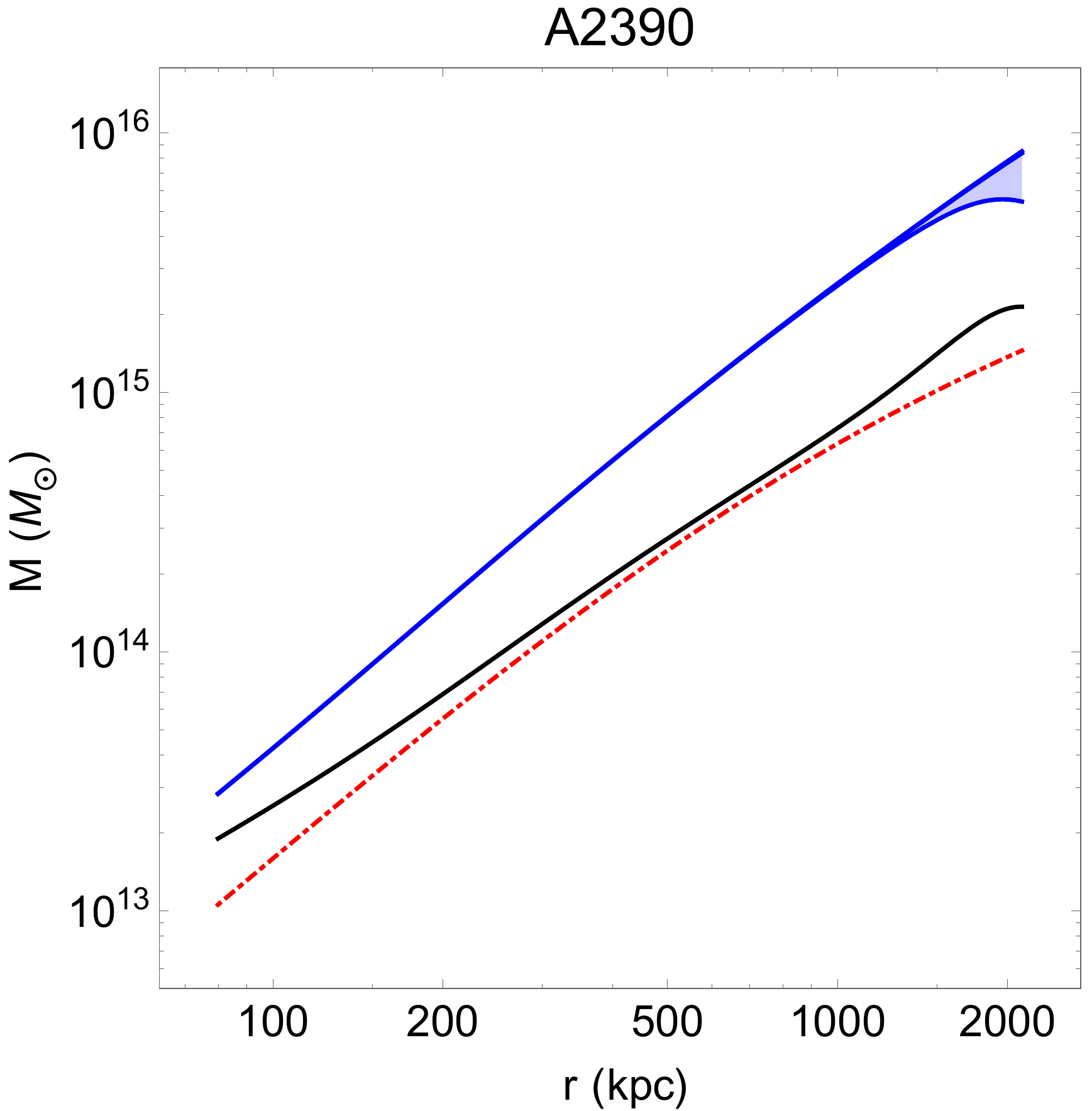}
\end{tabular}
\caption{Same as Figure \ref{MassPlotEMOND} for clusters A907 and A2390.}
\label{MassPlotEMOND6}
\end{figure*}

Using this recipe, it is therefore possible to calculate the effective enclosed mass predicted by EMOND for each cluster by which we mean Newtonian mass plus phantom mass,

\begin{equation}
M_{EMOND}(r) = \frac{r^{2} |\nabla \Phi_{EMOND}|}{G}
,\end{equation}

\noindent where $\nabla \Phi_{EMOND}$ is the gravitational acceleration given by Equation \ref{EMONDSpherical}. To accomplish this, we need to solve a first order differential equation for the potential $\Phi$.  This requires a boundary condition of the potential. For this we picked three values for the boundary, which we set at $r_{bound} = 2 r_{200}$ for each cluster.  The three values were $\Phi(r_{bound}) = \Phi_{NFW}(r_{bound})$,$\Phi(r_{bound}) = 0.5 \Phi_{NFW}(r_{bound}),$ and $\Phi(r_{bound}) = 1.5 \Phi_{NFW}(r_{bound})$. We did this for two reasons. First, we would expect that the potential can be approximated by the NFW profile and also we wanted to show how much the calculated mass is affected by the choice of boundary potential. In Figures \ref{MassPlotEMOND} to \ref{MassPlotEMOND6}, where we show the results for each cluster using the EMOND recipe, the shaded region highlights the dependence  of the boundary potential.

 The final aspect of the EMOND analysis that we should provide for completeness is proving that the EMOND $T_{2}$ term is in fact negligible. To do this, we can solve the right-hand side of Equation \ref{EMONDPoiss} by again using the NFW profile as an estimation of the total gravity in the clusters. We can then compare this result to the simplified Poisson equation, Equation \ref{EMONDPoiss2}, to see the effect that the $T_{2}$ term has on the result. Assuming the simple $\mu$ function (Equation \ref{simple}), as $F'(y) = \mu(\sqrt{y})$, $F(y)$ takes the form,

\begin{equation}
F(y) = -2\sqrt{y} + y + 2 \log(1 + \sqrt{y}).
\end{equation}

\noindent where  $y = |\nabla \Phi|^{2}/A_{0}(\Phi)^{2}$. Therefore,

\begin{equation}
y F'(y) - F(y) = \frac{2\sqrt{y} + y}{1 + \sqrt{y}}  - 2 \log(1 + \sqrt{y}).
\end{equation}

\noindent Assuming the simpler case where $A_{0}(\Phi)= a_{0}\exp(\Phi/\Phi_{0})$,

\begin{equation}
\frac{d (A_{0}(\Phi))^{2}}{d\Phi} = \frac{2a_{0}^{2}}{\Phi_{0}}\exp\left( \frac{2\Phi}{\Phi_{0}} \right).
\end{equation}

\begin{figure}
\includegraphics[scale=0.7]{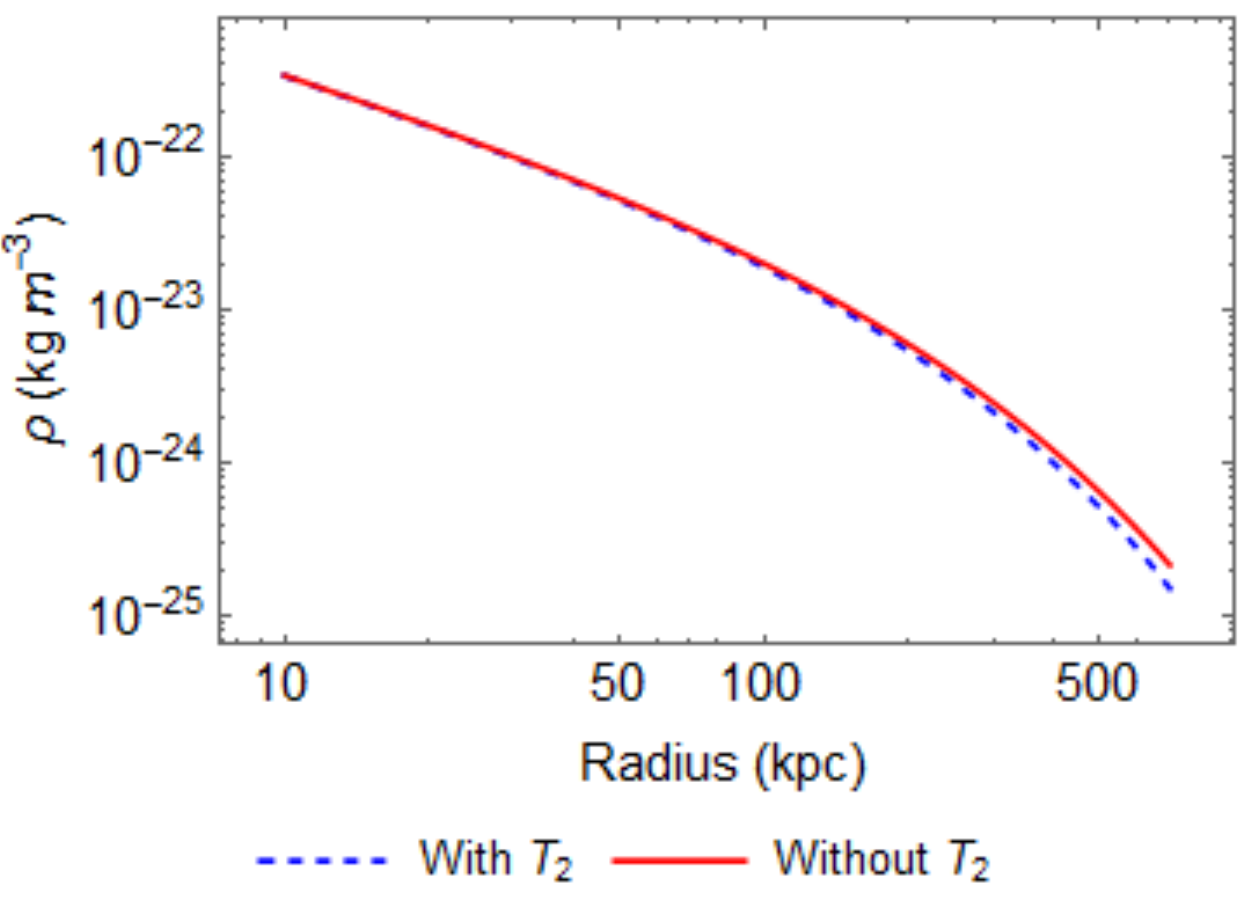}
\caption{ Plot showing the calculated density, predicted by EMOND for cluster A133. Blue dashed line shows the density with the inclusion of the $T_{2}$ term (Equation \ref{EMONDPoiss}) and red line shows  the density calculated from approximate Poisson equation (Equation \ref{EMONDPoiss2}). The lines are almost identical showing the $T_{2}$ term was indeed justifiably neglected. Note that the small differences in the outer regions of the cluster is an artifact because 
the asymptotic potential of NFW is compatible mathematically with EMOND only if the latter also allows the density dipping into negative at large radii.}
\label{T2Compute}
\end{figure}

We now have all the ingredients to numerically plot the $T_{2}$ term. To accomplish this, we show, for conciseness, the result for one cluster, A133, as an example but all clusters show similar results. In Figure \ref{T2Compute} we plot the full right-hand side (RHS) of Equation \ref{EMONDPoiss} and the RHS of \ref{EMONDPoiss2}. This is essentially a plot of the predicted Newtonian density profile from the EMOND formulation with the $T_{2}$ term included and neglected. We see that the plot shows that the results are almost identical and thus we were justified in neglecting the $T_{2}$ term in our previous analysis. The values deviate from each other in the outer radii of the cluster. This is because the NFW profile is not a perfect solution of the EMOND equation and the predicted Newtonian density falls to zero and eventually negative. The difference in the two lines is just a result of the small delay between the two terms driving the density to this zero point. 

We conclude from this analysis that EMOND has had mixed success as a MOND generalization to explain the missing mass in galaxy clusters. We therefore look for an alternative in the following section.

\section{An empirical alternate formulation explaining the mass discrepancy}\label{AMONDSection}

\begin{figure}
\includegraphics[scale=0.6]{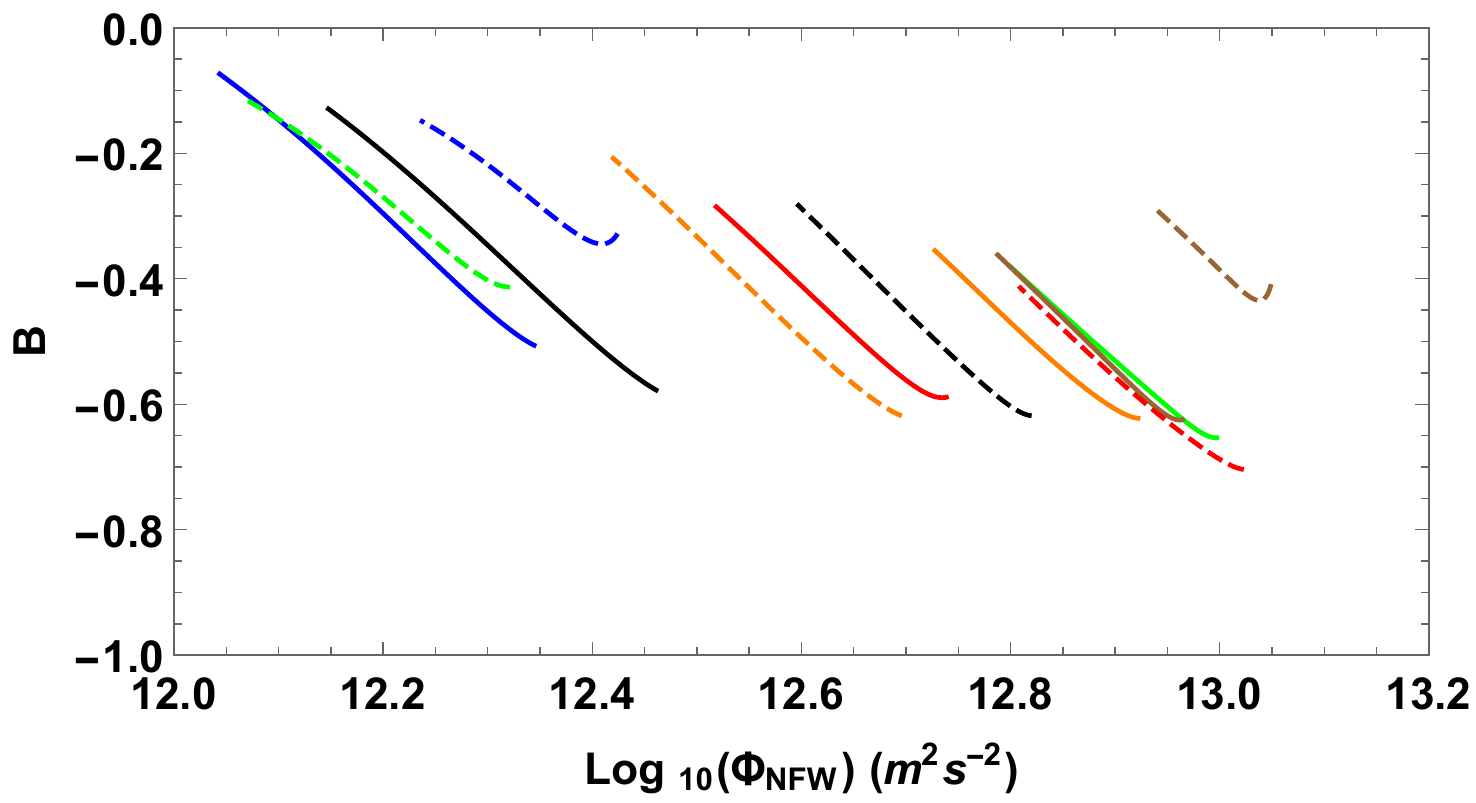}
\caption{Plot showing the quantity B (Equation \ref{Bplot}) vs. NFW gravitational potential for each cluster.    By approximating the total cluster gravity as that of the NFW halo, we illustrate that 
any correction term B to the regular MOND formula, while having a trend with the potential of each cluster, cannot describe all clusters simultaneously.}
\label{Bplot}
\end{figure}

 Despite the mixed success of EMOND in answering the question of missing mass in galaxy clusters, the trend of missing mass versus gravitational potential seems to linger. Also the idea of having $a_{0}$ as a non-constant function may not sit comfortably with the MOND community. This prompted us to look for a different formalism that could unify galaxies and galaxy clusters with one law. We thought it would be a good idea to determine the residual in the MOND formula, which we will call B, to determine if there is some common theme throughout the clusters. The MOND residual is found by rearranging the MOND formula such that,

\begin{equation}\label{Beqn}
B = \frac{|\nabla \Phi_{N}|}{|\nabla \Phi|}- \mu\left( \frac{|\nabla\Phi|}{a_{0}} \right),
\end{equation}

\noindent
which is just moving everything in Equation \ref{MONDPoiss} to one side, assuming spherical symmetry. The hope here was that we could find a function for B that is 0 in galaxies, thus preserving regular MOND, but non-zero in galaxy clusters. We can determine whether there is a trend if we plot the value of B for each cluster in our sample,  assuming that the total gravitational acceleration is approximated by the best-fit NFW profile as a function of total gravitational potential of each cluster; this gravitational potential is also approximated by the gravitational potential of the NFW profile.  With this approach, we attempt to determine whether there is an additive component to the original MOND formula that can boost the gravity inside clusters, whilst preserving regular MOND in galaxies. We plot this result in Figure \ref{Bplot}.

One may notice that although there is not a single function that could fit all these clusters, there seems to be some regularity to the plot.
To determine whether the result in Figure \ref{Bplot} can be improved,  we tried modifying our Equation for B (Equation \ref{Beqn}) slightly such that,

\begin{equation}\label{B2eqn}
B_{2} = \frac{|\nabla \Phi_{N}|}{|\nabla \Phi|}- \mu\left( \frac{|\nabla\Phi|}{a_{0}} +\frac{\Phi}{\Phi_{0}} \right).
\end{equation}

\begin{figure}
\includegraphics[scale=0.6]{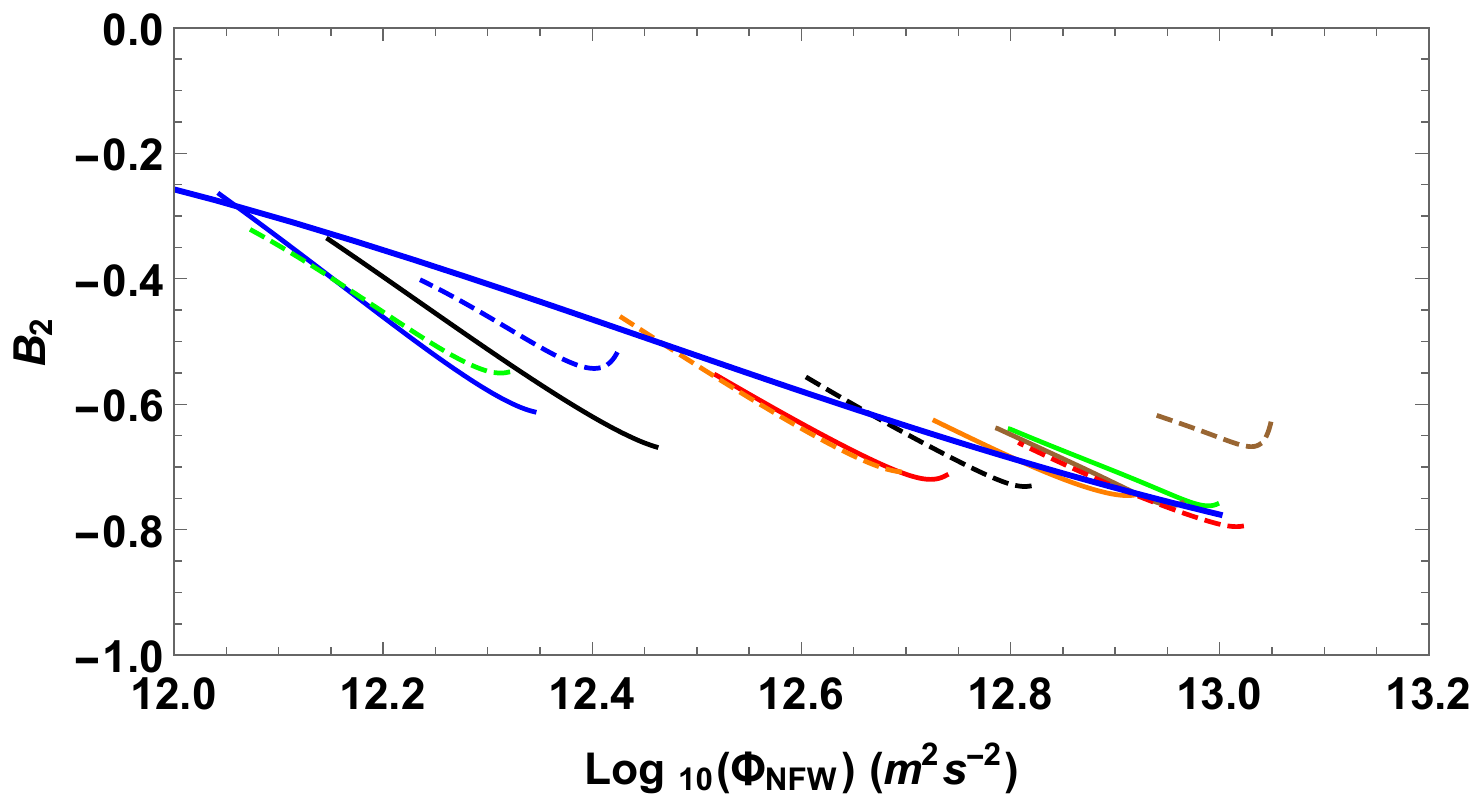}
\caption{Plot showing the value $B_{2}$ (Equation \ref{B2eqn}) as a function of NFW gravitational potential for each cluster (thin lines). We overplot a thin blue line that shows $\mu(\Phi/\Phi_{0})$ for $\Phi_{0}=-1700000^{2}$ m$^{2}$s$^{-2}$. Note all clusters lie fairly close to this line, with some discrepancy
due to our adopted NFW profile not representing the total potential.
This $B_2$ correction to MOND allows the MOND interpolation function to run with the potential as well as the acceleration.}
\label{B2plot}
\end{figure}

The result of plotting $B_{2}$ against gravitational potential is shown in Figure \ref{B2plot}. This yielded  a very tight function of gravitational potential for each cluster. Also, one should notice that the absolute value of $B_{2}$ runs between 0 and 1, which is identical to the behaviour of the MOND interpolating function. Although the trend is very tight, there are some offsets. This may be because we approximated the gravitational potential to be the NFW potential. This assumption may not hold.

\subsection{Summarizing and refining the formulae}

Analysing the data seems to show a possible alternative formulation of MOND that could account for the mass discrepancy in  this sample of clusters. The modified empirical relation in this case would be

\begin{equation}
\frac{|\nabla \Phi_{N}|}{|\nabla \Phi|} -  \mu \left( \frac{|\nabla \Phi|}{a_{0}} + \frac{\Phi}{\Phi_{0}} \right) = B_{2}(\Phi)
,\end{equation}

where the gravitational potential, $\Phi$, and scale potential, $\Phi_{0}$, are negative quantities. As we have seen in Figure \ref{B2plot}, $B_{2}$ looks like a MOND interpolation function, $B_{2}(\Phi) = \mu(\Phi/\Phi_{0}(\Phi))$. Implementing this, our new modified MOND function is written as

\begin{equation}\label{AMOND1}
\nabla \Phi_{N} = \left[ \mu \left( \frac{|\nabla \Phi|}{a_{0}} + \frac{\Phi}{\Phi_{0}} \right) - \mu\left( \frac{\Phi}{\Phi_{0}}\right)  \right] \nabla \Phi,\end{equation}

\noindent
where $\Phi_{0}$ is our scale potential analogous to the MOND acceleration scale $a_{0}$.

Equation \ref{AMOND1} is a relation based on the MOND paradigm, which in the future can hopefully lead to a gravity theory that can explain away the mass discrepancy in galaxy clusters whilst preserving the dynamics of galaxies.

We have yet to mention the value for the scale potential $\Phi_{0}$. For our results, we chose a scale potential of $|\Phi_{0}| = 1700000^{2}\text{m}^{2}\text{s}^{-2}$. We do not attempt to perform a rigorous Monte Carlo search over the parameter space in this work as  we feel that to do this properly would require some independent tests of the equations such as gravitational lensing and Milky Way data to ensure our parameter does not cause contention with local observations. This is best left for further work.

\subsection{Comparison to MOND}
\begin{figure}
\includegraphics[scale=0.8]{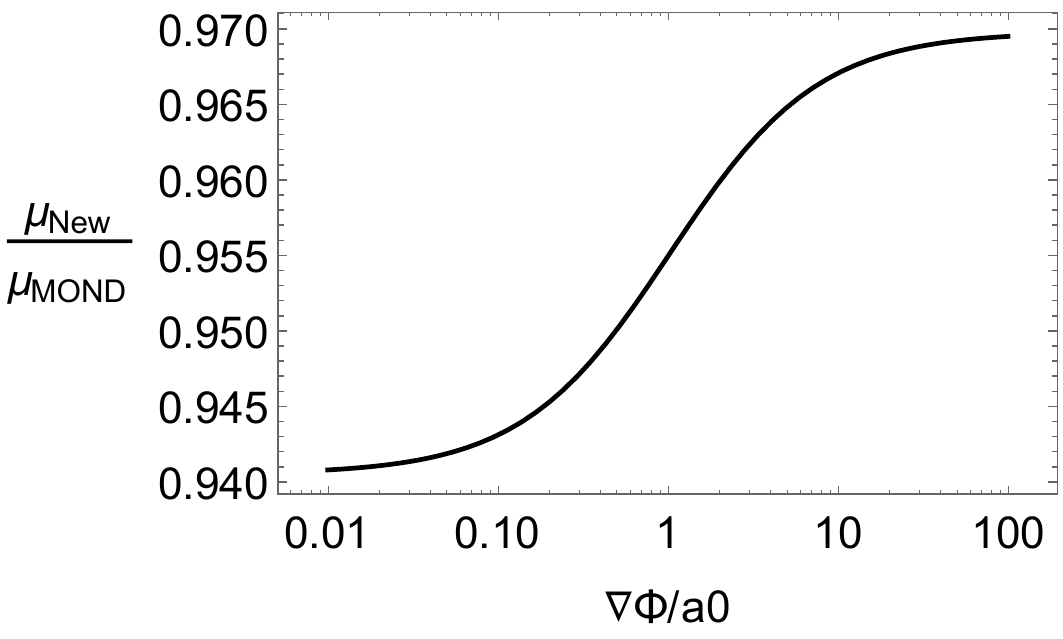}
\caption{Plot showing how the MOND interpolation function and the interpolation function of our relation (Equations \ref{AMONDmu} and \ref{MONDmu}) compare for a low potential object ($\Phi=\Phi_{ext}$) for different values of acceleration. We can see little variation, which is ideal for preserving galaxy physics with this new relation.}
\label{mucompare}
\end{figure}

We do however show a plot showing how the interpolation function of our relation compares with that of regular MOND (Figure \ref{mucompare}). Here,

\begin{equation}\label{AMONDmu}
\mu_{New} =  \left[ \mu \left( \frac{|\nabla \Phi|}{a_{0}} + \frac{\Phi}{\Phi_{0}} \right) - \mu\left( \frac{\Phi}{\Phi_{0}}\right)  \right]
\end{equation} 

\noindent
and 

\begin{equation}\label{MONDmu}
\mu_{MOND} = \mu\left( \frac{|\nabla \Phi|}{a_{0}} \right).
\end{equation}

\begin{figure}
\includegraphics[scale=0.6]{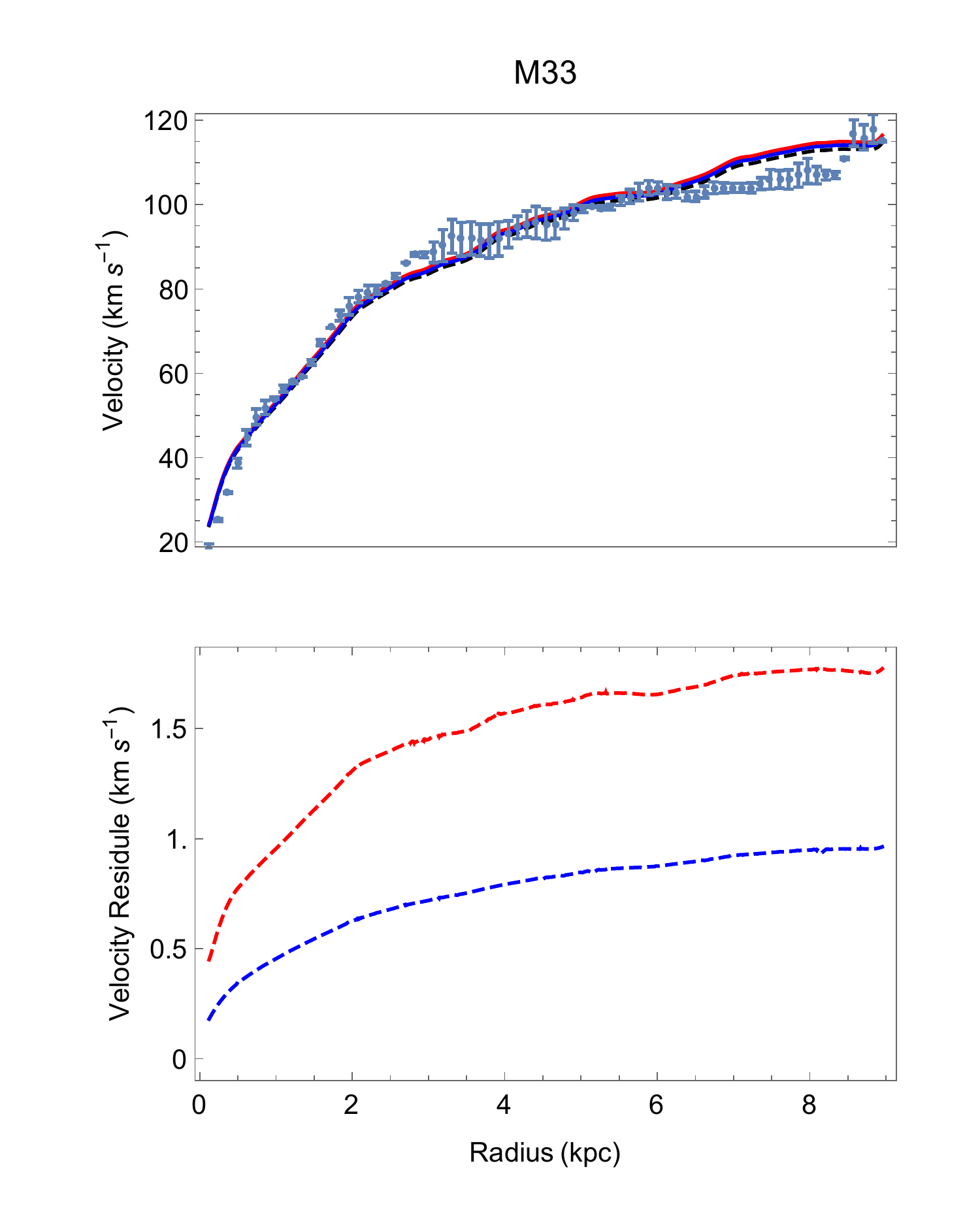}
\caption{Top plot showing the rotation curve of  M33. Blue points are the data with error bars, the black dashed line is the regular MOND fit, the red line is the new relation, and the blue line is EMOND. In the bottom plot we also show the residual between MOND, the new relation, and EMOND. Note the difference between these rotation curves are actually tiny ($\sim 1$km/s). Here we adopted a background potential of $10^{-6}c^2$ given in \cite{EMOND}), and a stellar mass to light of 0.722.}
\label{AMONDrotcurves1}
\end{figure}

\begin{figure}
\includegraphics[scale=0.6]{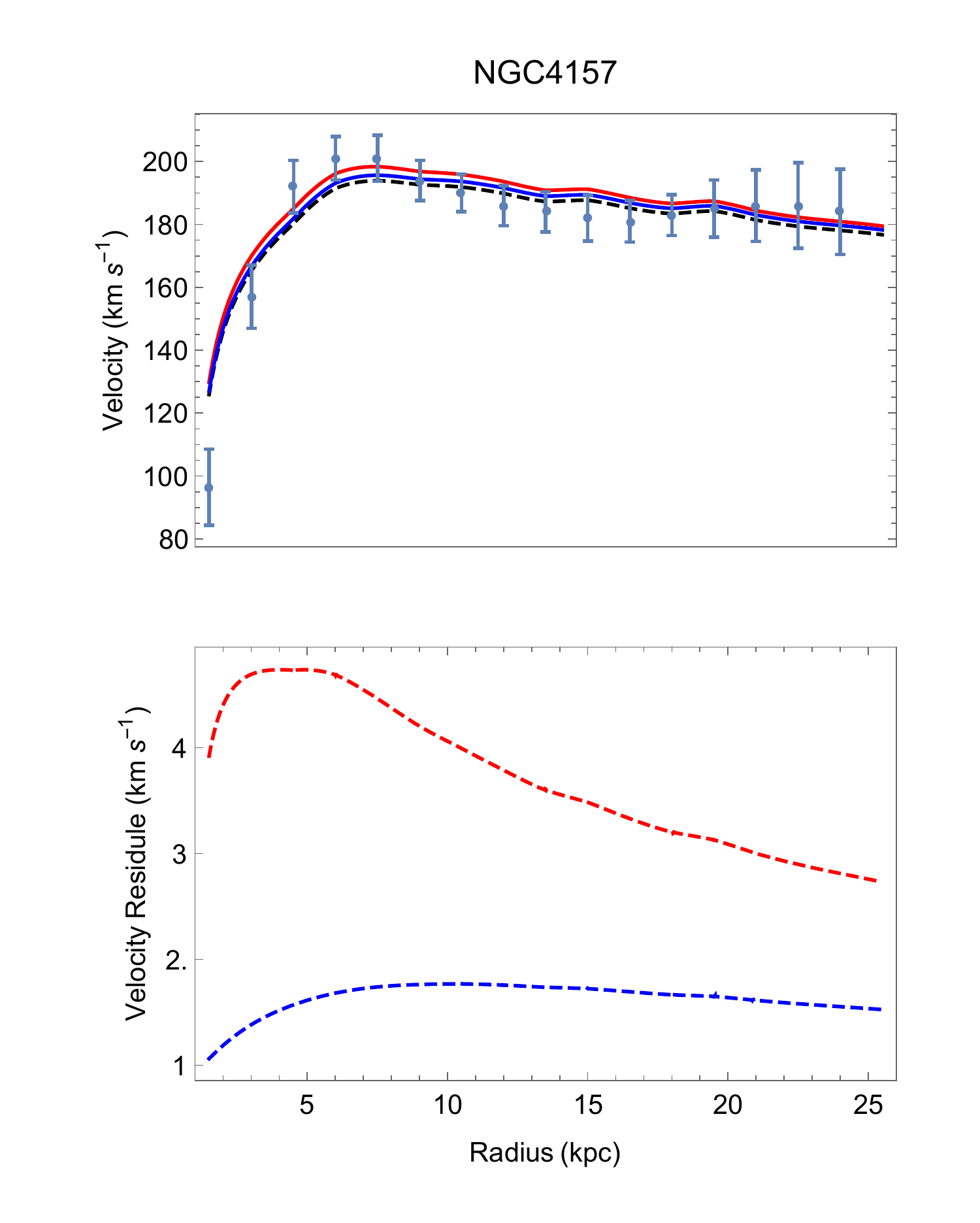}
\caption{Same as Figure \ref{AMONDrotcurves1} for galaxy NGC4157. We use a stellar mass to light of 1.69 for each case.}
\label{AMONDrotcurves2}
\end{figure}

We compare Equations \ref{AMONDmu} and \ref{MONDmu} in Figure \ref{mucompare}, assuming that the potential $\Phi= \Phi_{ext}$ as defined in Section \ref{ExtPot} for different values of acceleration $a$ ranging from $0.01 a_{0} - 100 a_{0}$. We also show rotation curves for two galaxies, M33 and NGC4157, using the new relation and EMOND (Figures \ref{AMONDrotcurves1} and \ref{AMONDrotcurves2}). These rotation curves have been used in \citet{m33rotcurve} for M33 and \citet{NGC4157rotcurve} for NGC 4157. We overplot the MOND prediction for comparison. There is little deviation from MOND for these two examples. There may be some degeneracy between fitting the scale potential for a galaxy and the stellar mass to light. We do not discuss this point here, but mention  it as a possible avenue for future work. We have used a fixed background potential for both galaxy cases whereas in practice, this may not be correct. Ideally one would make a larger scale simulation encompassing a large area of space and work out the external potential (and field) for each galaxy. This was mentioned in \citep{decliningcurves} who used the external field to explain declining rotation curves. They also mentioned that this was now possible thanks to the MOND patch in RAMSES, Phantom of RAMSES \citep{RAMSES} and \citep{POR}. This would require more work to incorporate for example EMOND as the gravity solver would need to be altered to allow varying $a_{0}$ values.

\section{Comparing the new relation to the NFW fits}\label{AMONDNFWSection}

In this section, we provide plots of the dynamical mass, the mass predicted by invoking the new relation and $\Lambda$CDM predictions for the 12 clusters outlined in Section \ref{SampleSection}. The dynamical mass is calculated via Equation \ref{DynamicalMass}, the mass predicted from the new relation,

\begin{equation}
M_{New}(r) = \frac{r^{2} \nabla \Phi_{New}}{G}
\end{equation}

\noindent
where $\nabla\Phi_{New}$ is calculated via Equation \ref{AMOND1} and the $\Lambda$CDM mass from,

\begin{equation}\label{MassLCDM}
M_{\Lambda CDM}(r) = M_{gas}(r) + M_{BCG} + M_{NFW}(r).
\end{equation}

The NFW parameters used in Equation \ref{MassLCDM} are taken from \citep{sample} and are displayed here in Table \ref{concentration} in which the NFW mass is defined as

\begin{equation}
M_{NFW}(r) = 4 \pi \rho_{s} r_{s}^{3} \left[ \ln \left( 1 + \frac{r}{r_{s}} \right) - \left( 1 + \frac{r_{s}}{r} \right)^{-1} \right]
,\end{equation} 

\noindent
where $\rho_{s}$ and $r_{s}$ are the scale potential, and radius of the NFW profile $r_{s}$ is defined as $r_{s} = r_{500}/c_{500}$, where the values for $r_{500}$ and $c_{500}$ (the NFW concentration parameter) can be found in Tables \ref{r500} and \ref{concentration}, respectively. The value $\rho_{s}$ is calculated from the equation,

\begin{equation}\label{rhoscale}
\rho_{s} = \frac{500}{3} \left( \frac{ r_{500}}{r_{s}} \right)^{3}\frac{\rho_{crit}}{\log \left[ 1 + \frac{r_{500}}{r_{s}} \right]- \left[ 1 + \frac{r_{s}}{r_{500}} \right]^{-1}}
,\end{equation}

\noindent where $\rho_{crit}$ is the critical density.

\begin{table}
\centering
\caption{NFW concentration parameters at radius $r_{500}$. NFW profiles can be determined by combining these with the $r_{500}$ parameters given in Table \ref{r500}, the values for $r_{s}$ and the value for $\rho_{s}$ given in Equation \ref{rhoscale}.}
\label{concentration}
\begin{tabular}{|l|l|}
\hline
Cluster & $c_{500}$ \\
\hline
A133    & 3.18                  \\
A262    & 3.54                  \\
A478    & 3.57                  \\
A1413   & 2.93                  \\
A1795   & 3.21                  \\
A1991   & 4.32                  \\
A2029   & 4.04                  \\
RXJ1159 & 1.7                   \\
MKW4    & 2.54                  \\
A383    & 4.32                  \\
A907    & 3.48                  \\
A2390   & 1.66                                  \\          
\hline     
\end{tabular}
\end{table}

\begin{figure*}
\begin{tabular}{ccc}
\includegraphics[scale=0.5]{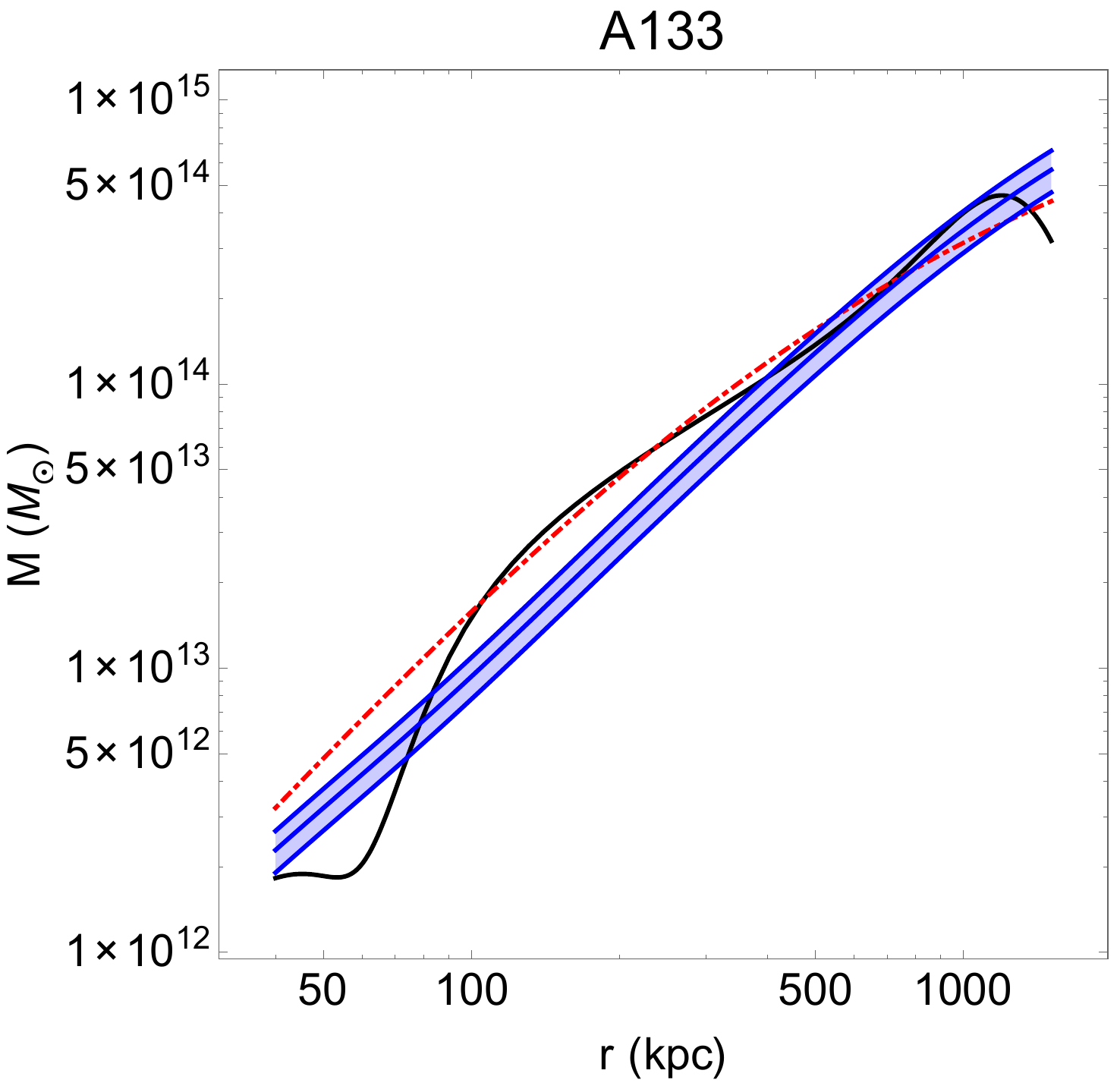} & \includegraphics[scale=0.5]{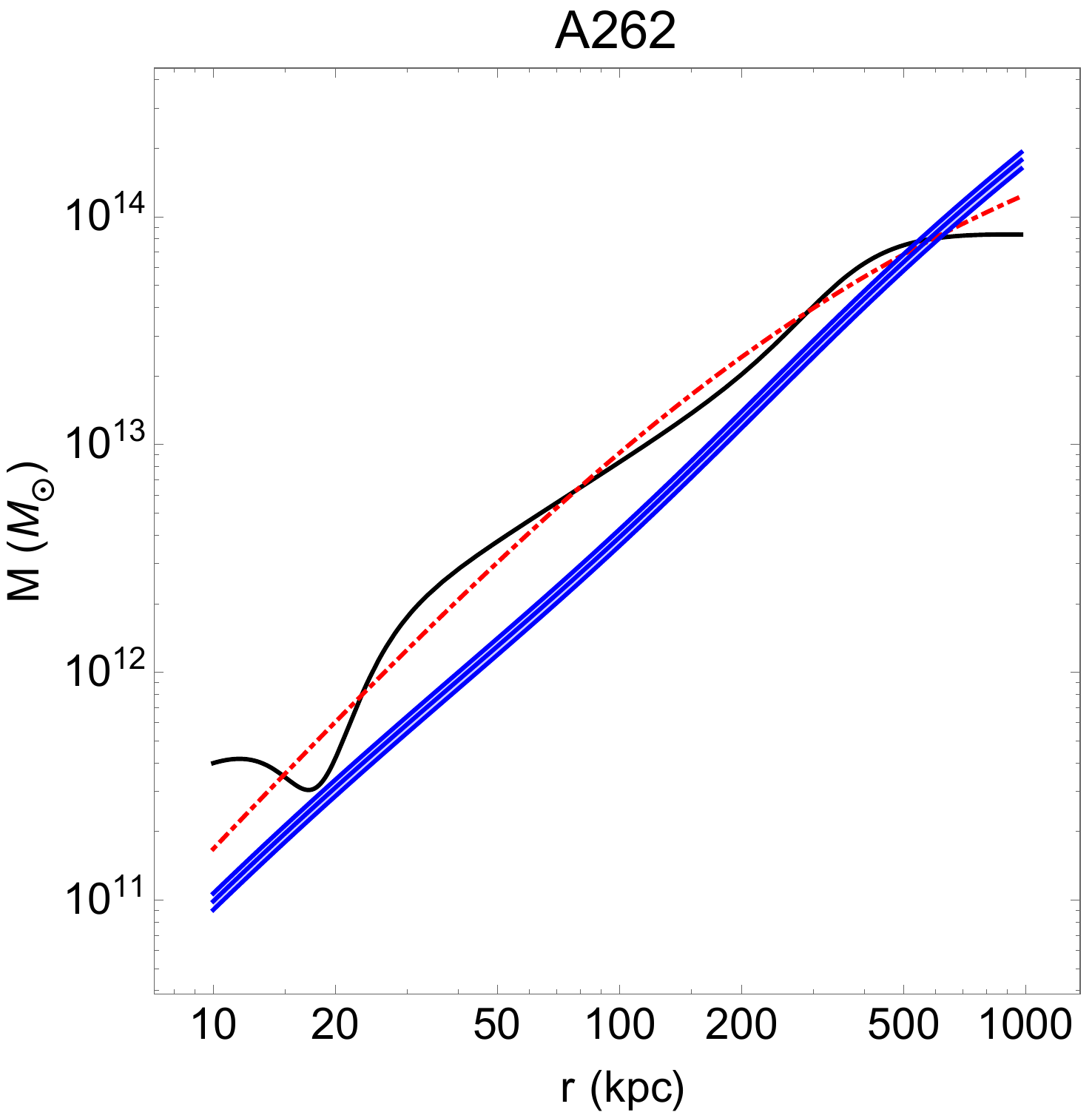}
\end{tabular}
\caption{Plot showing mass profiles for the new relation for different values for the boundary potential (blue shaded region). Also the plot shows the NFW (red dotted line) and dynamical masses (black solid line) for clusters A133 and A262. }
\label{MassPlotAMOND}
\end{figure*}

\begin{figure*}
\begin{tabular}{ccc}
\includegraphics[scale=0.5]{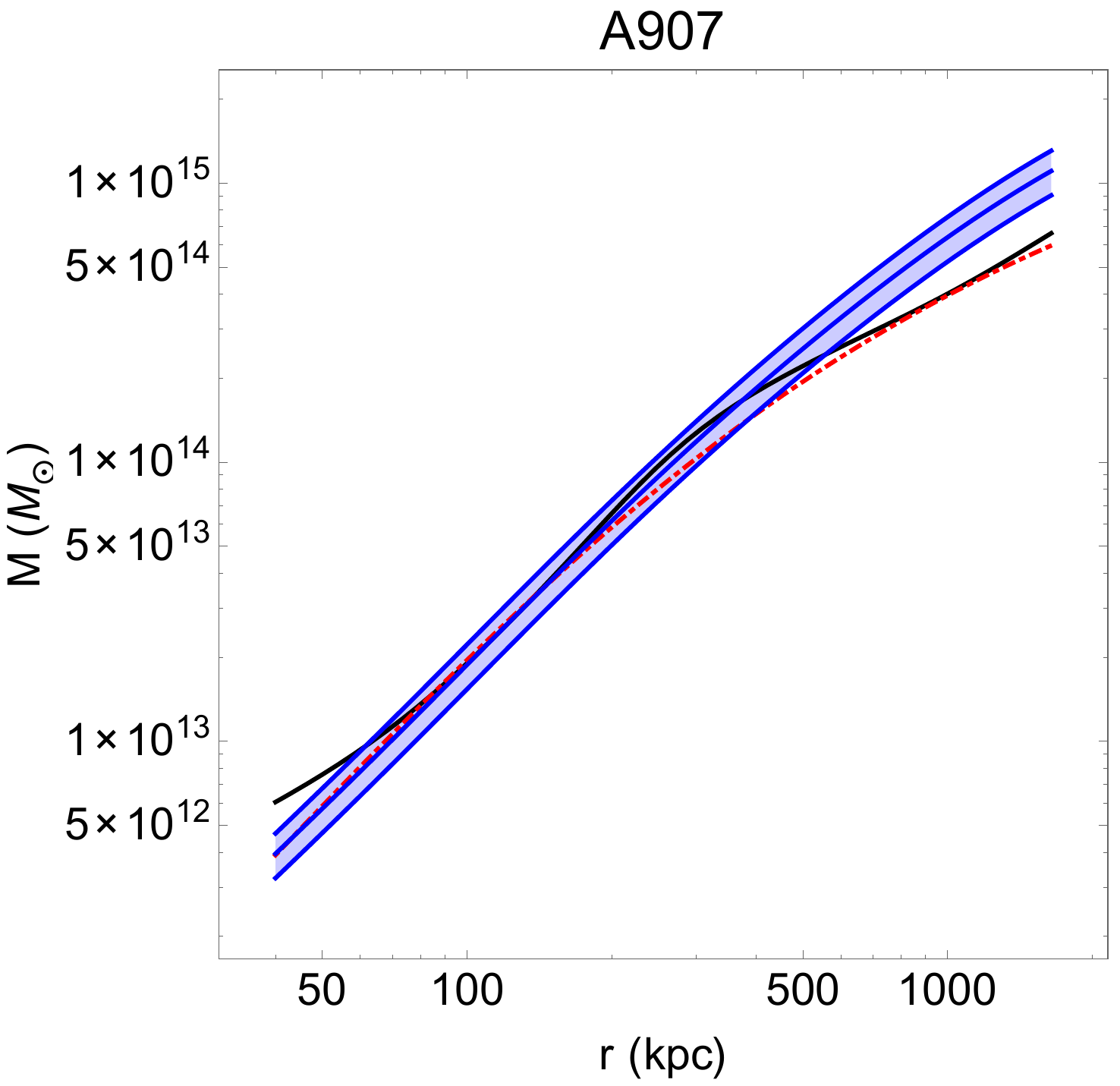} & \includegraphics[scale=0.5]{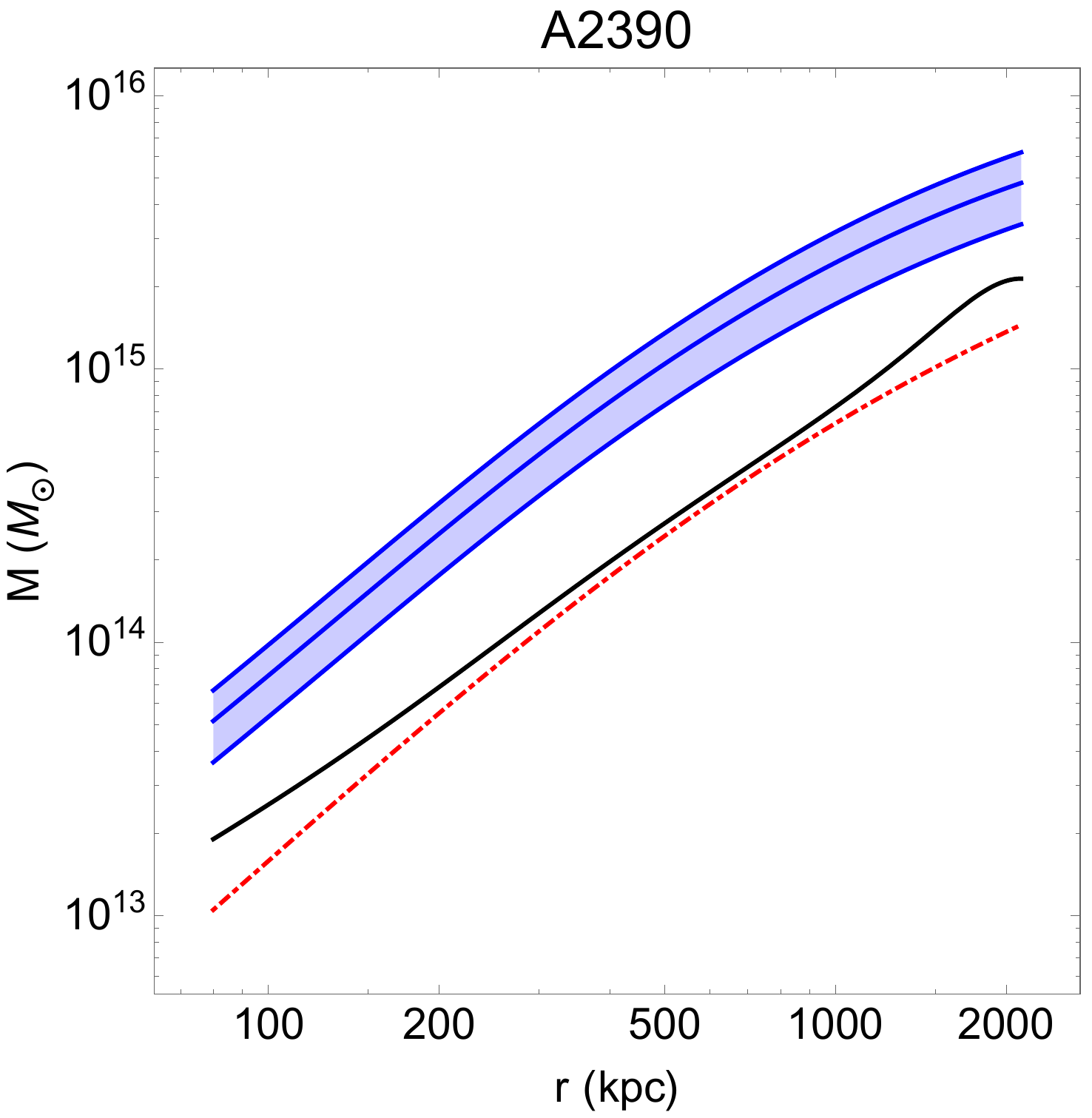}
\end{tabular}
\caption{Same as Figure \ref{MassPlotAMOND} for clusters A907 and A2390.}
\label{MassPlot6}
\end{figure*}

We can see from the dynamical mass plots (Figures \ref{MassPlotAMOND}-\ref{MassPlot6}) for each cluster that on average the new relation can provide a reasonable boost, or at least a reasonable match to the NFW profile, for the sample. The most noticeable outlier from our new relation is A2390, which has a very poor fit. A2390 is the largest cluster and thus has the largest gravitational potential. This new relation, in this current from, has a problem in very high gravitational environments (see Section \ref{AMONDLimitationSection}). Perhaps a refinement of the formalism in these high potential regions may solve the problem of A2390. We also acknowledge that a rigorous error analysis has not been performed and thus it is hard to gauge how well this new relation has faired in predicting cluster masses. 

Also, we do not truncate the gas density in our calculation. It is unrealistic that the gas density extends to $r_{200}$. As we do not have the necessary data to constrain this, we also leave this as an open issue.

\section{Limitations of the new relation and required testing of the formalism}\label{AMONDLimitationSection}

Although the new relation can recover the mass of the cluster to reasonable precision, there is a severe flaw in the empirical law. Both the gravitational potential and acceleration are large near stars and black
holes. Therefore we have $\frac{|\nabla \Phi|}{a_{0}} + \frac{\Phi}{\Phi_{0}}>>1$ and $ \frac{\Phi}{\Phi_{0}}>>1$. This results in  

\begin{equation}
\left[ \mu \left( \frac{|\nabla \Phi|}{a_{0}} + \frac{\Phi}{\Phi_{0}} \right) - \mu\left( \frac{\Phi}{\Phi_{0}}\right)  \right] \rightarrow 1 - 1 \rightarrow 0.
\end{equation} 

\noindent
and thus Equation \ref{AMOND1} becomes,

\begin{equation}
\nabla \Phi = \nabla \Phi_{N}/0 \rightarrow \infty.
\end{equation}

Therefore, although this new formula can cover cluster and galaxy scales, it inevitably fails on the stellar scale as the predicted gravity tends towards infinity. This problem is inherent to the structure of Equation \ref{AMOND1}, which has one interpolation function subtracting another. Perhaps when constructing a Lagrangian based formulation of the law, there will arise some extra terms that could counter this problem. This is an issue which is not addressed here, but in further work.

Despite these two issues, it is still possible to further test this law with regard to gravitational lensing and by looking at the dynamics of objects within clusters, which will be affected by the external potential effect. Possible candidates for the latter could be rotation curves for intercluster galaxies or the mass estimates for the newly discovered ultra-diffuse galaxies \citep{ultradiffusemass}, all of which are beyond the scope of this work.

Finally, more rigorous data analysis should be conducted to determine whether better interpolation functions can be found and a parameter search could be implemented to determine better model parameters.

\section{Discussion and conclusions}\label{Conclusion}

In this work, we look at the possibility that the missing mass in galaxy clusters, which is still present in MOND, may be attributed to a modified gravity law that is a generalization of MOND. Beginning with an overview of a previous study concerning this area, we review the equations of EMOND and apply the formalism to a sample of galaxy clusters. We show that EMOND in its current form has partial success in explaining the clusters.

We then go on to use the data to determine if there is any empirical relation that could explain the missing mass in galaxy clusters. We find that there is such an expression, which works relatively well for this sample of clusters. The new formula has the ability to account for the missing mass in clusters by modifying the MOND formula in such a way that the gravity is boosted in regions of high gravitational potential whilst preserving the regular MOND formula in regions of weak gravitational potential.

The main issue with this modified equation however, is that in regions of high acceleration and
high potential, such as near a star or black hole, the enhancement of gravity calculated with the
formula tends to infinity and is thus in contradiction with observation.  
A thorough investigation is also needed to refine the interpolation functions and the new scale potential value
by expanding the data set of clusters and invoking 
new tests of the law, such as gravitational lensing and dynamics of objects that lie within galaxy
clusters.  Furthermore, the empirical equations here should have a Lagrangian formulation, which would also have 
implications for cosmology. While it is difficult 
to predict the applications of our model to cosmology partly because of the lack of a covariant
Lagrangian, the effect of a lower MOND acceleration scale in galaxies compared to galaxy
clusters in our model could mean that the structure formation is suppressed in galaxy scale
compared to cluster scale. This might skew the relative abundances of bound systems and
suppress the luminosity functions at the lower end. 

It is very difficult at this stage to speculate on any implications for cosmology and structure formation with the new relation we describe as it still has theoretical challenges. We can however make some comments with regards to EMOND. Firstly, there is a known connection between dark energy and MOND: mainly, $(8 a_{0})^{2}/(8\pi G) \approx \Lambda$. If, like in EMOND, $a_{0}$ is increased, and there is a link between the acceleration  scale factor and dark energy, then EMOND may affect the dark energy contribution and would also imply that the contribution is uneven rather than constant. This could cause differences in late time expansion and also the angular size of the CMB peaks compared to regular MOND. We would also expect that lensing would be enhanced if a covariant version were formulated. The main test for EMOND would be the lensing signature of the bullet cluster. \cite{EMOND} discuss lensing of a bullet cluster-like object in their original work. They find that it is possible to create phantom dark matter-like effects offset from the baryons in EMOND. Originally, \cite{EMOND} only boosted the acceleration of the scale factor by a factor of $\approx 6,$ where as we make a much larger boost of a factor $\approx 50$. We should therefore expect larger effects than in the original work. Clearly, to make a stronger case, we would actually have to make a non-spherical model to see if EMOND is in fact consistent with the bullet cluster (and train wreck cluster). With regards to galaxy formation, we might expect galaxies to form differently within clusters as the effective $a_{0}$ is larger. We mentioned earlier that the nature of the UDGs might be compatible with an EMOND-like theory; perhaps simulations in EMOND could produce these objects. We might also expect that galaxies form differently near the centre of a cluster compared to at the edge as $A_{0}(\Phi)$ is very different in the centre than at the cluster edge. One prediction might be that UDGs at the cluster edge should show a less severe dynamical M/L ratio than those closer to the centre.

All these are beyond the scope of our work here, which is to show that there could be an empirical
gravity relation that can, without actually invoking dark matter, account for the missing mass in galaxy clusters

\section*{ACKNOWLEDGEMENTS}

We would like to thank Benoit Famaey for useful discussions.  AOH is supported by Science and Technologies Funding Council (STFC) studentship (Grant code: 1-APAA-STFC12). We would also like to thank the referee for very useful comments and Anne-Marie Weijmans for proofreading the draft.

\begin{figure*}
\begin{tabular}{ccc}
\includegraphics[scale=0.5]{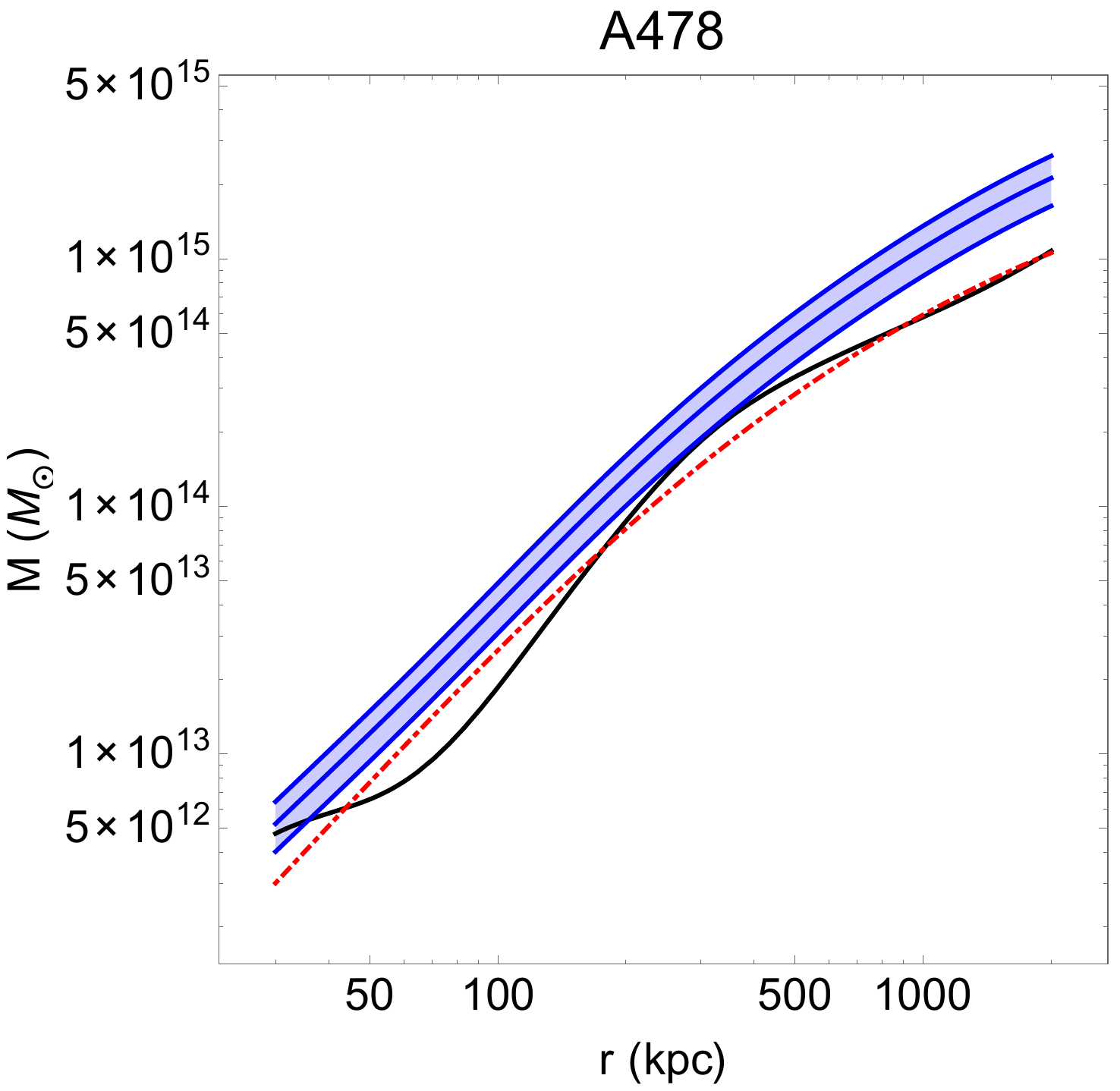} & \includegraphics[scale=0.5]{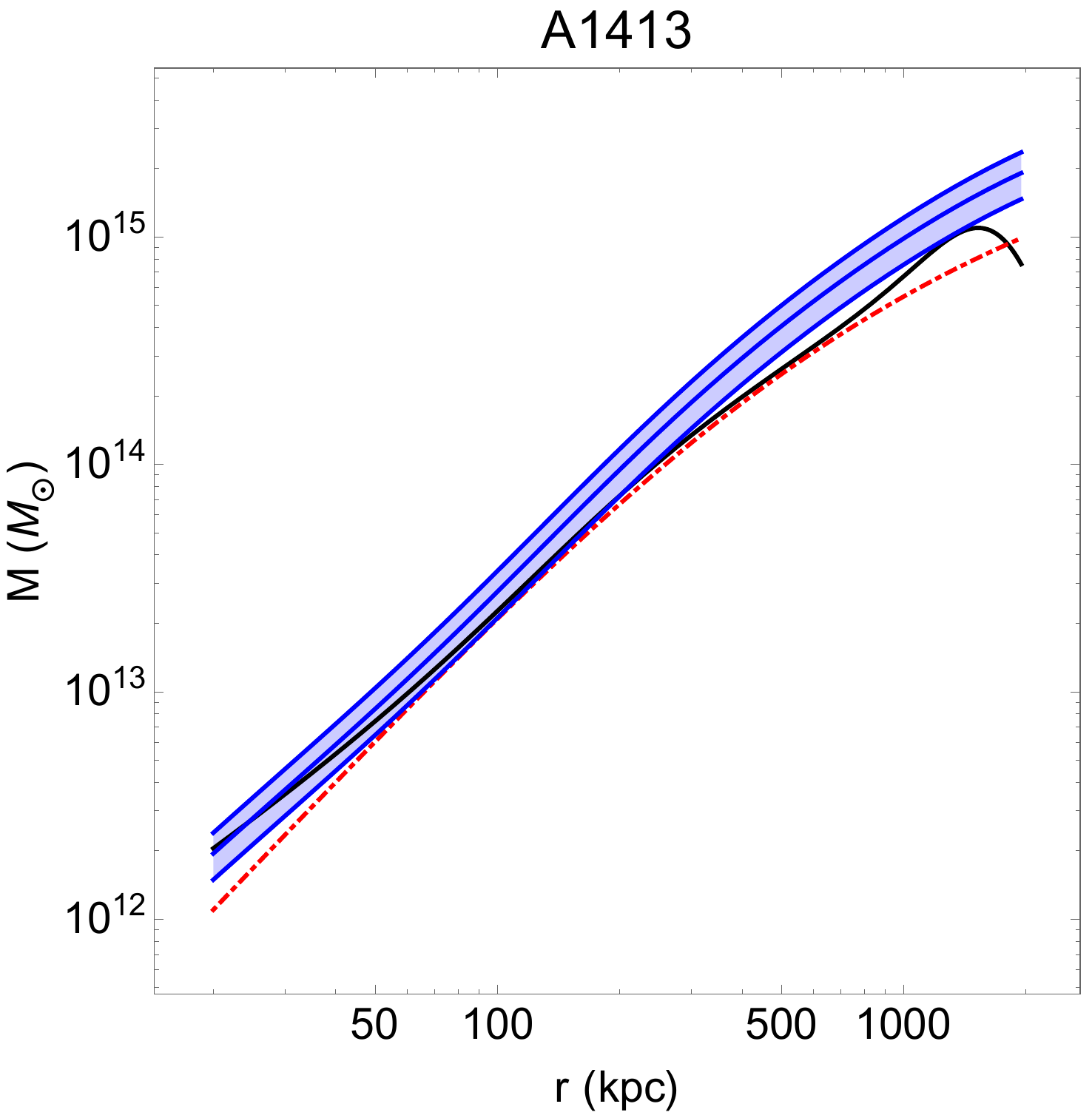}
\end{tabular}
\caption{Same as Figure \ref{MassPlotAMOND} for clusters A478 and A1413.}
\label{}
\end{figure*}

\begin{figure*}
\begin{tabular}{ccc}
\includegraphics[scale=0.5]{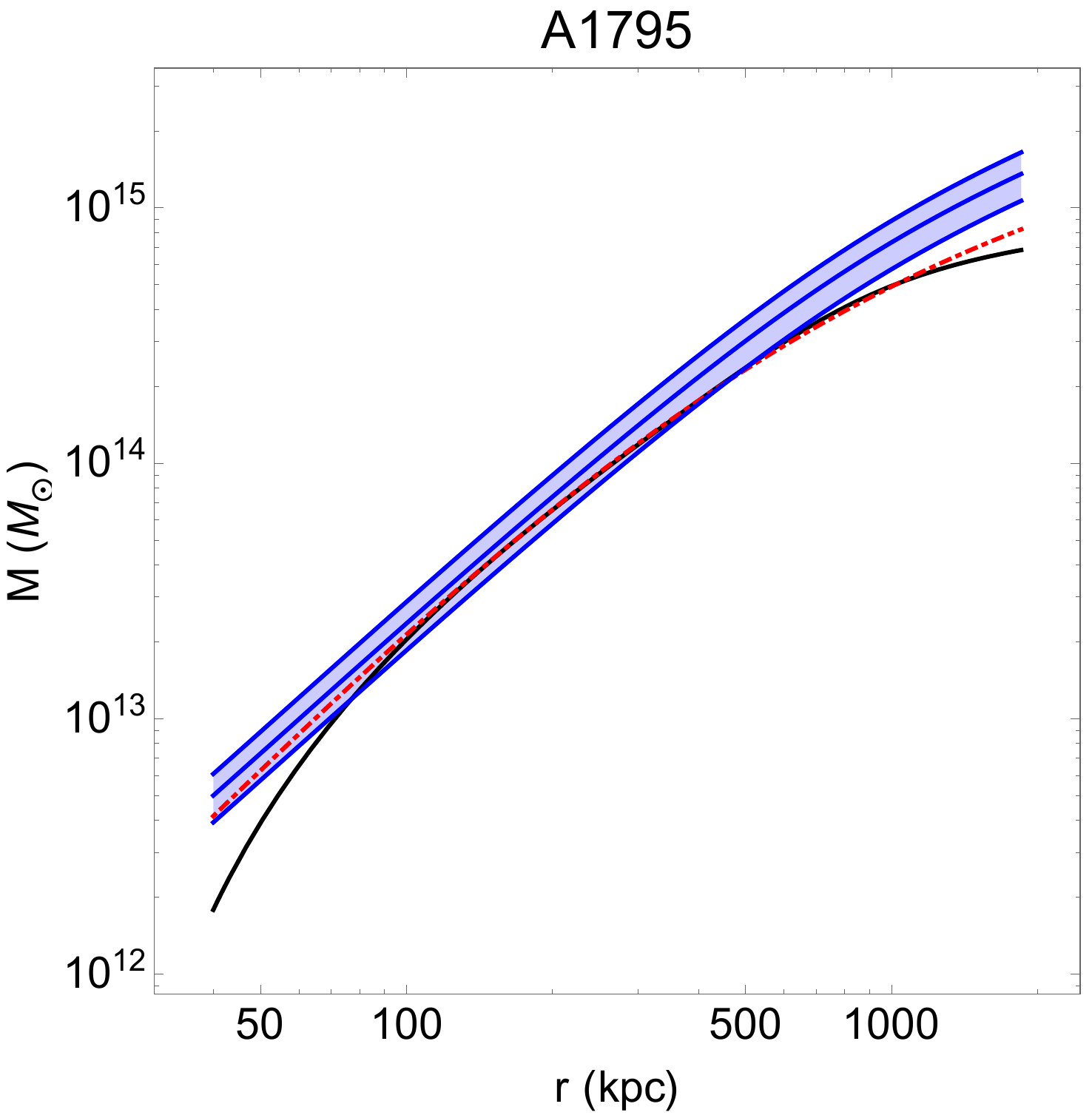} & \includegraphics[scale=0.5]{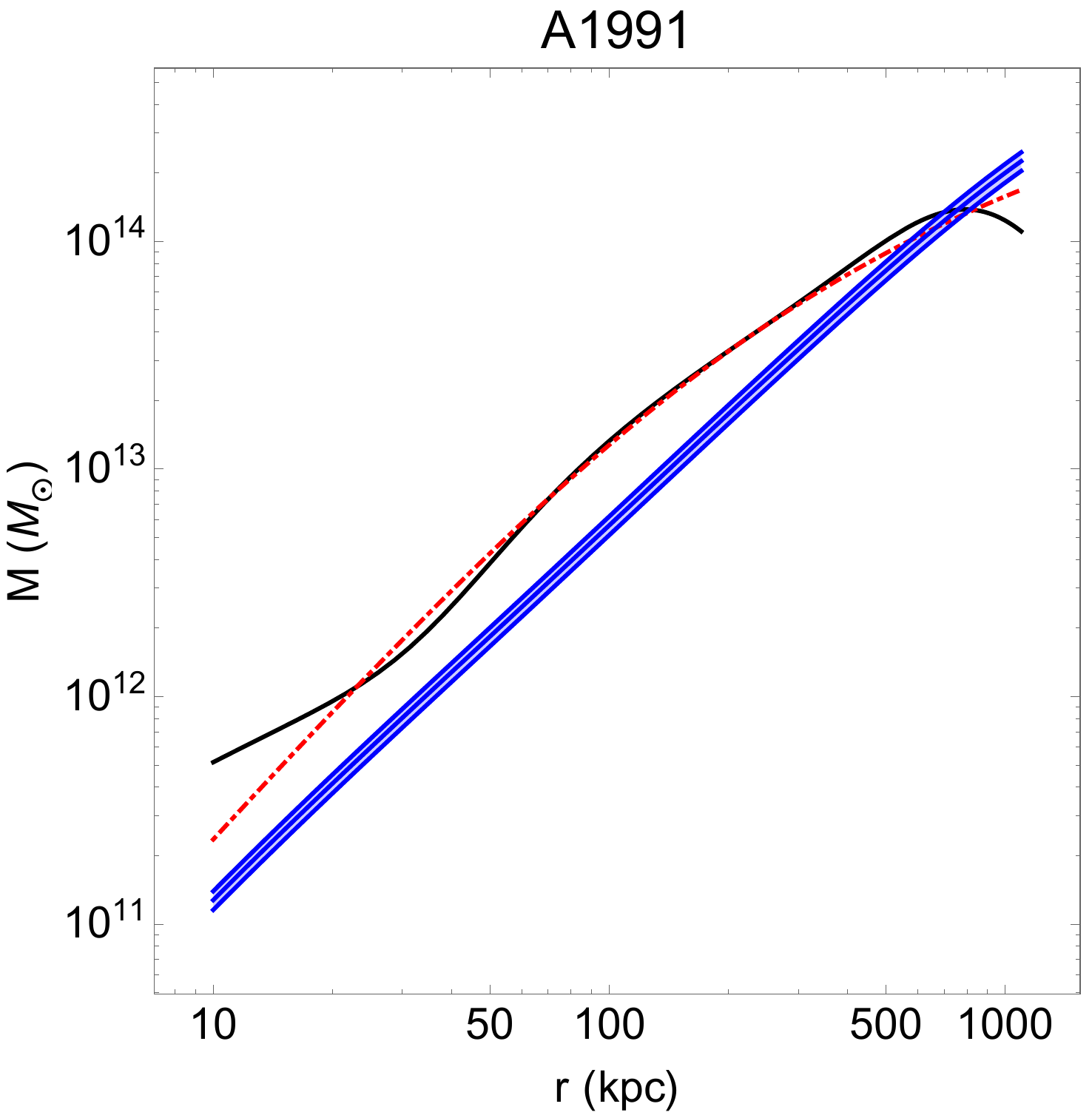}
\end{tabular}
\caption{Same as Figure \ref{MassPlotAMOND} for clusters A1795 and A1991.}
\label{}
\end{figure*}

\begin{figure*}
\begin{tabular}{ccc}
\includegraphics[scale=0.5]{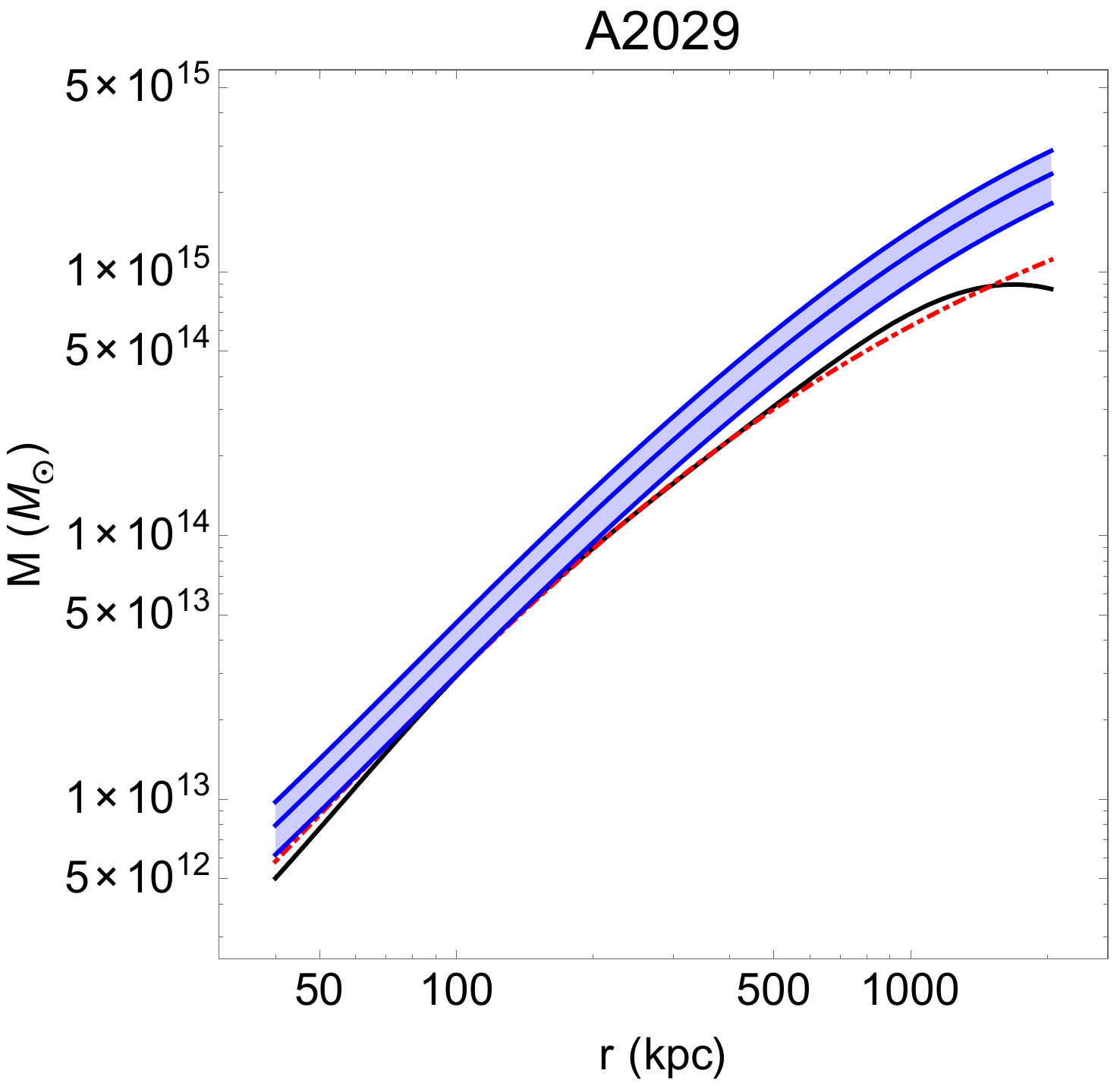} & \includegraphics[scale=0.5]{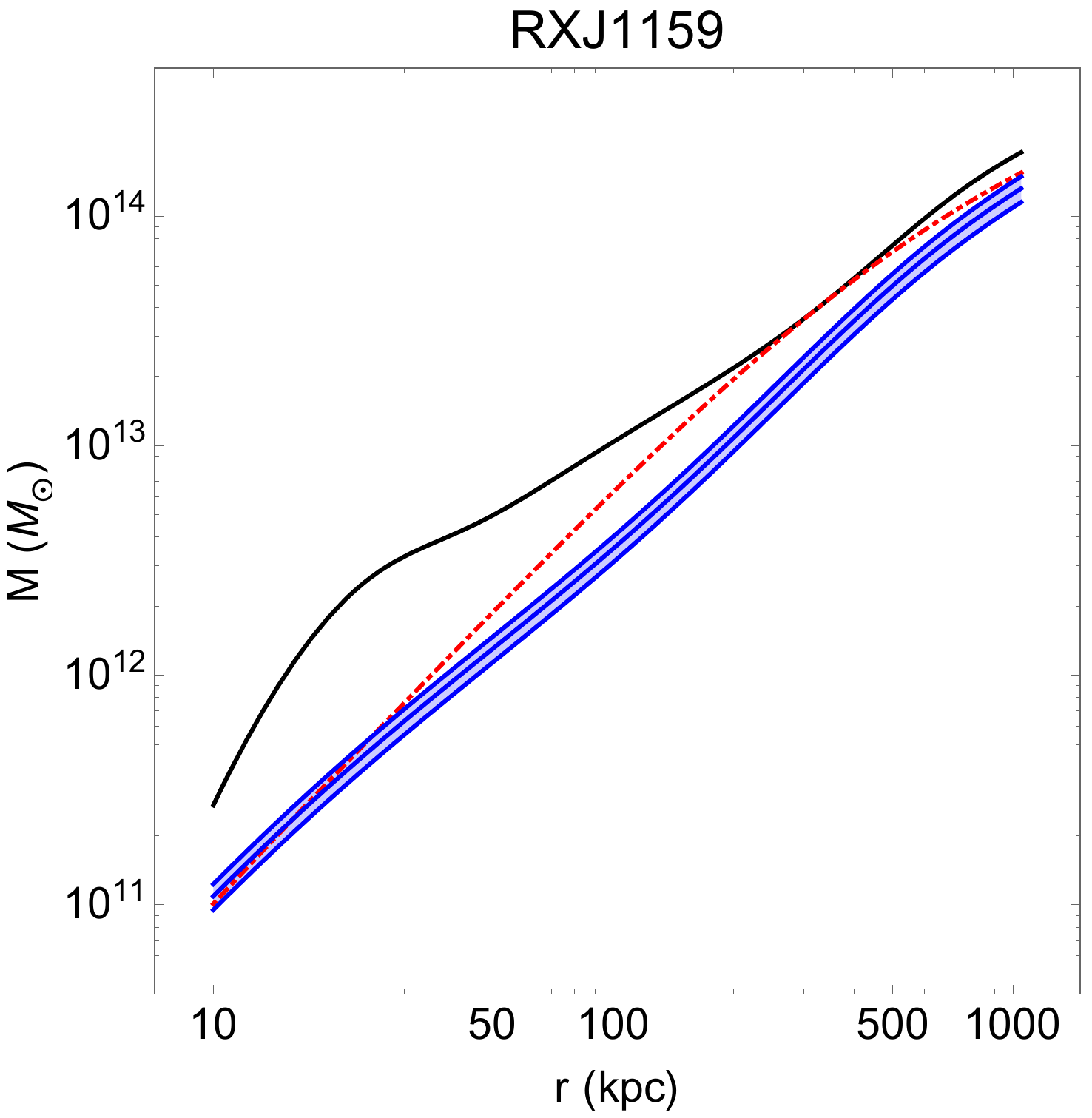}
\end{tabular}
\caption{Same as Figure \ref{MassPlotAMOND} for clusters A2029 and RXJ1159.}
\label{}
\end{figure*}

\begin{figure*}
\begin{tabular}{ccc}
\includegraphics[scale=0.5]{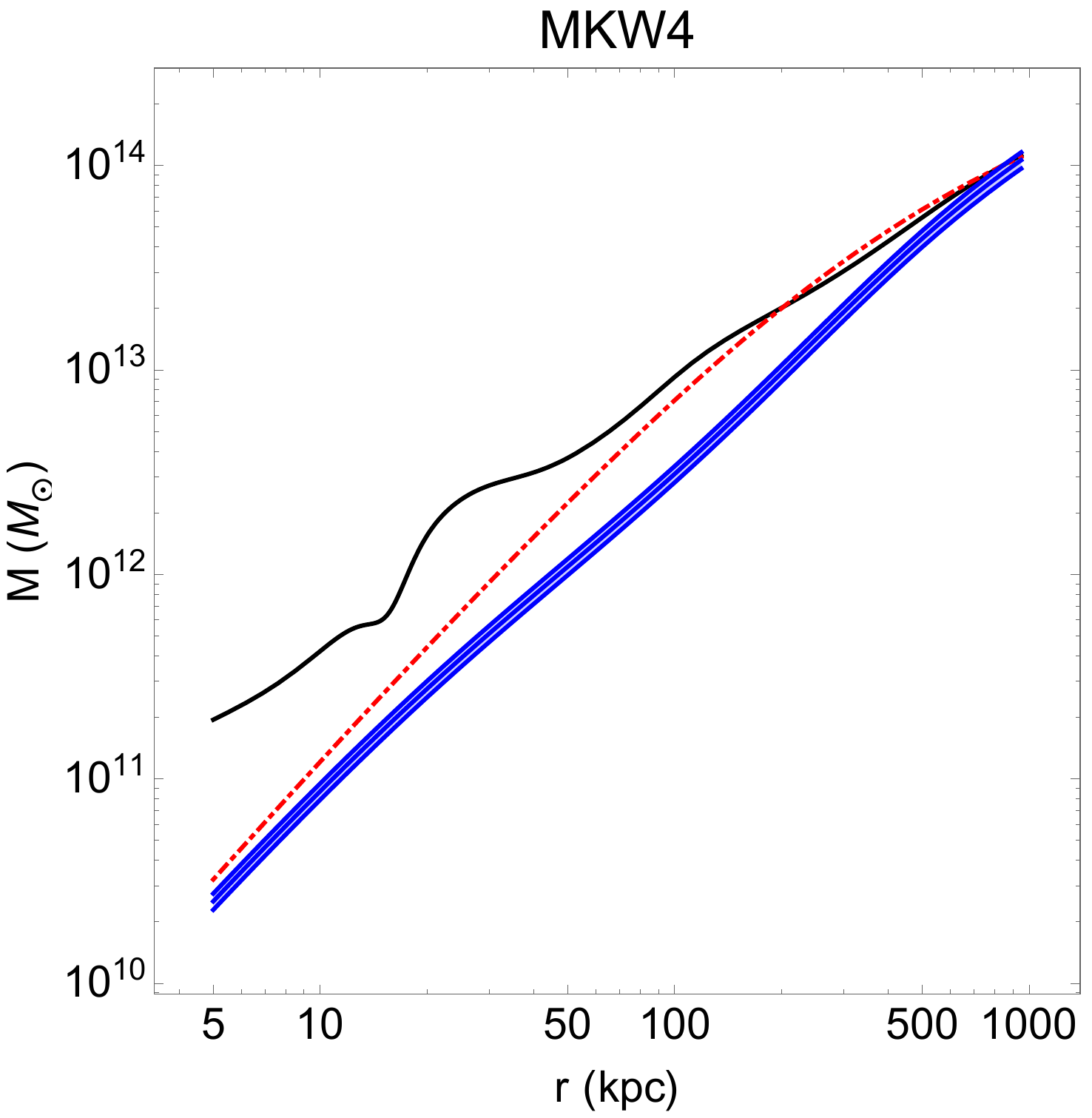} & \includegraphics[scale=0.5]{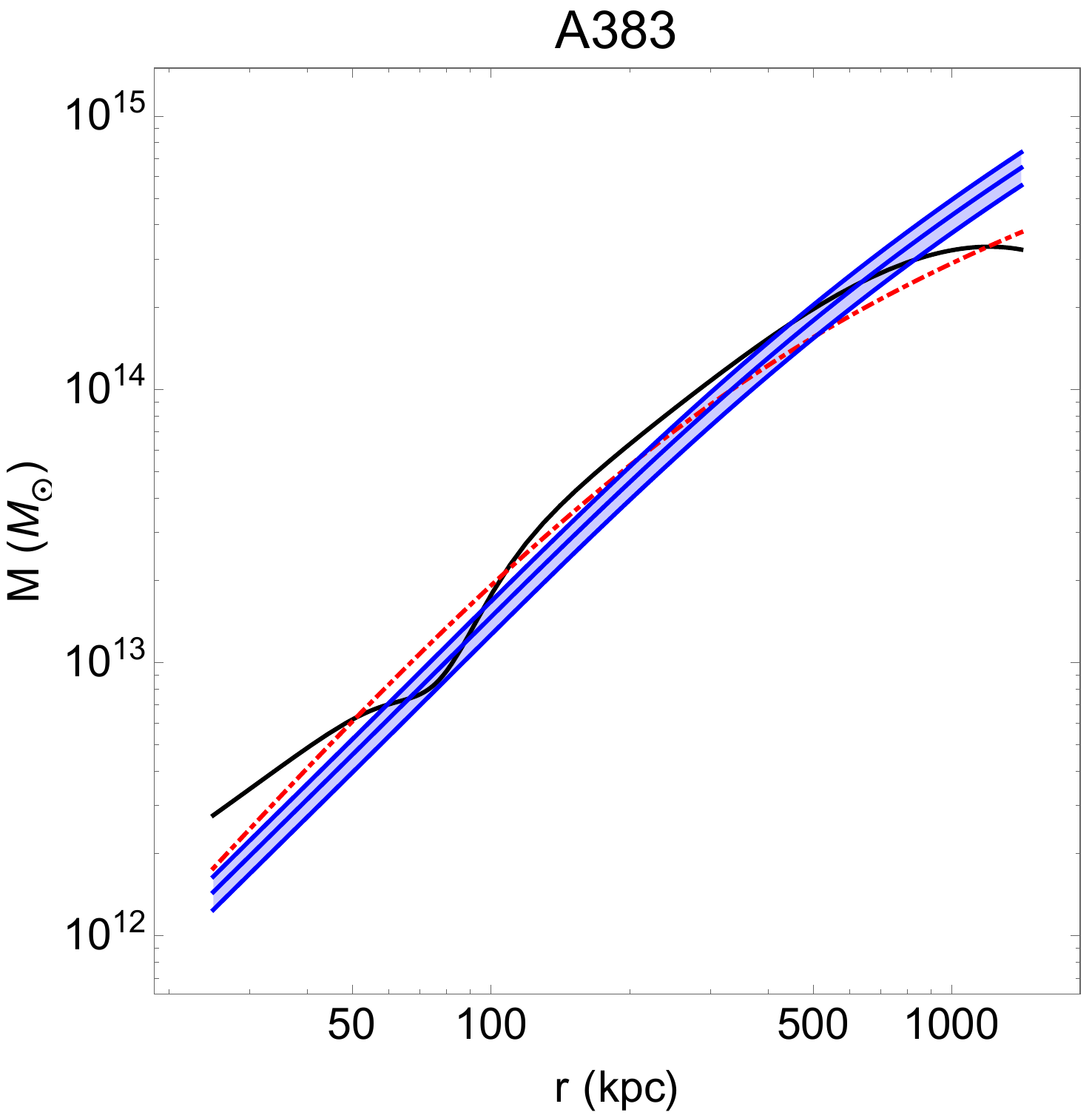}
\end{tabular}
\caption{Same as Figure \ref{MassPlotAMOND} for clusters MKW4 and A383.}
\label{}
\end{figure*}

\begin{figure*}
\begin{tabular}{ccc}
\includegraphics[scale=0.5]{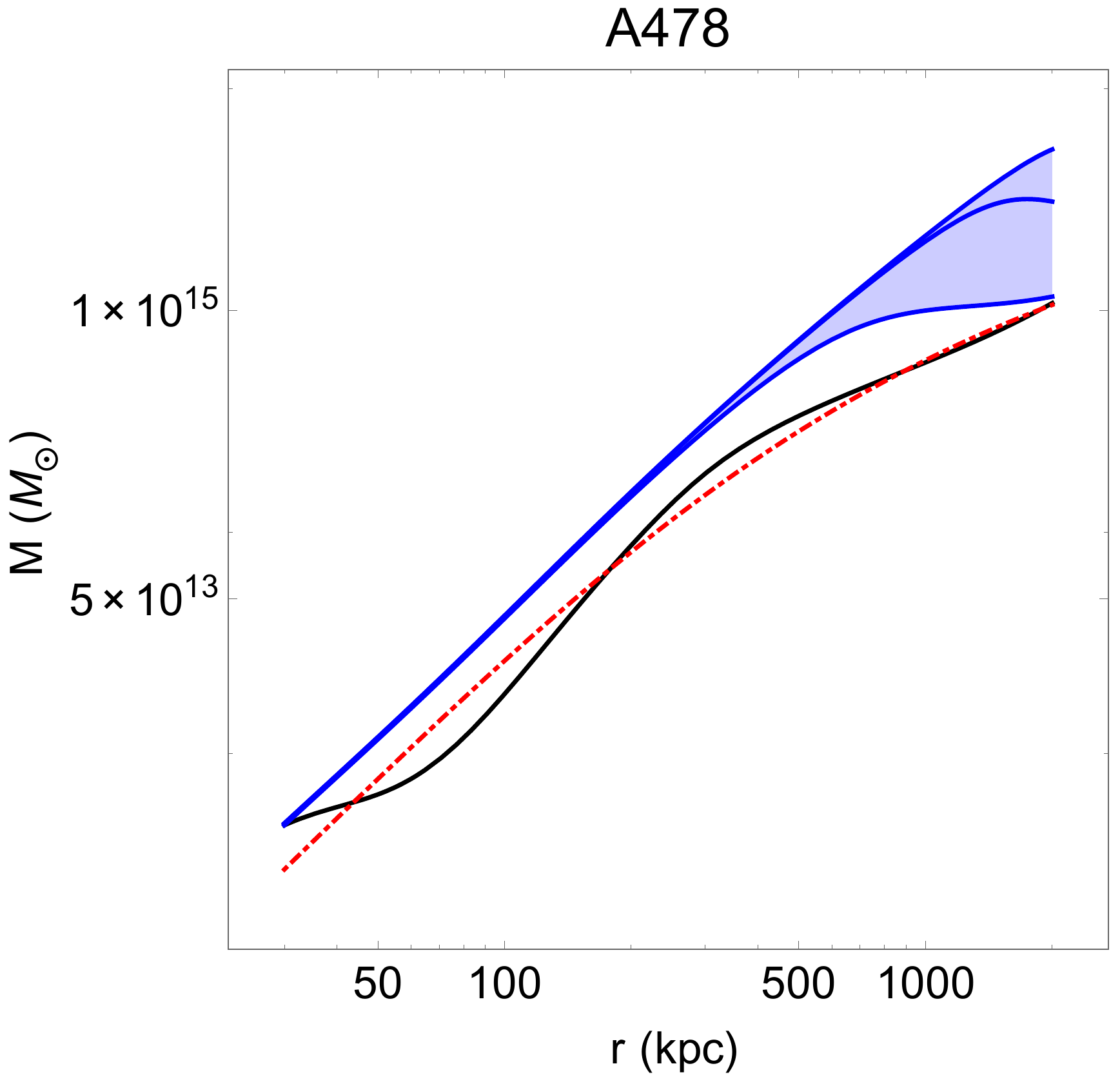} & \includegraphics[scale=0.5]{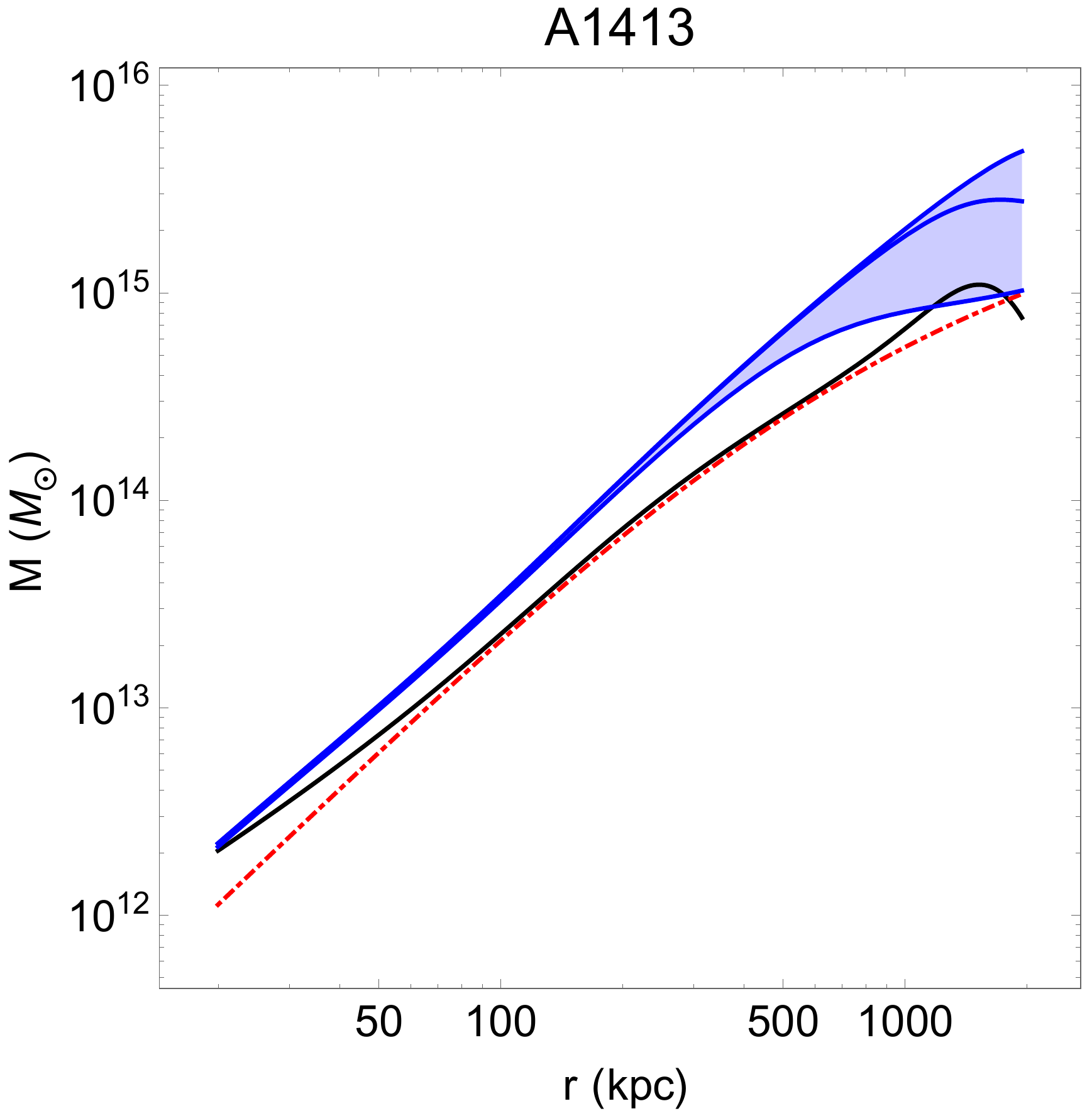}
\end{tabular}
\caption{Same as Figure \ref{MassPlotEMOND} for clusters A478 and A1413.}
\label{}
\end{figure*}

\begin{figure*}
\begin{tabular}{ccc}
\includegraphics[scale=0.5]{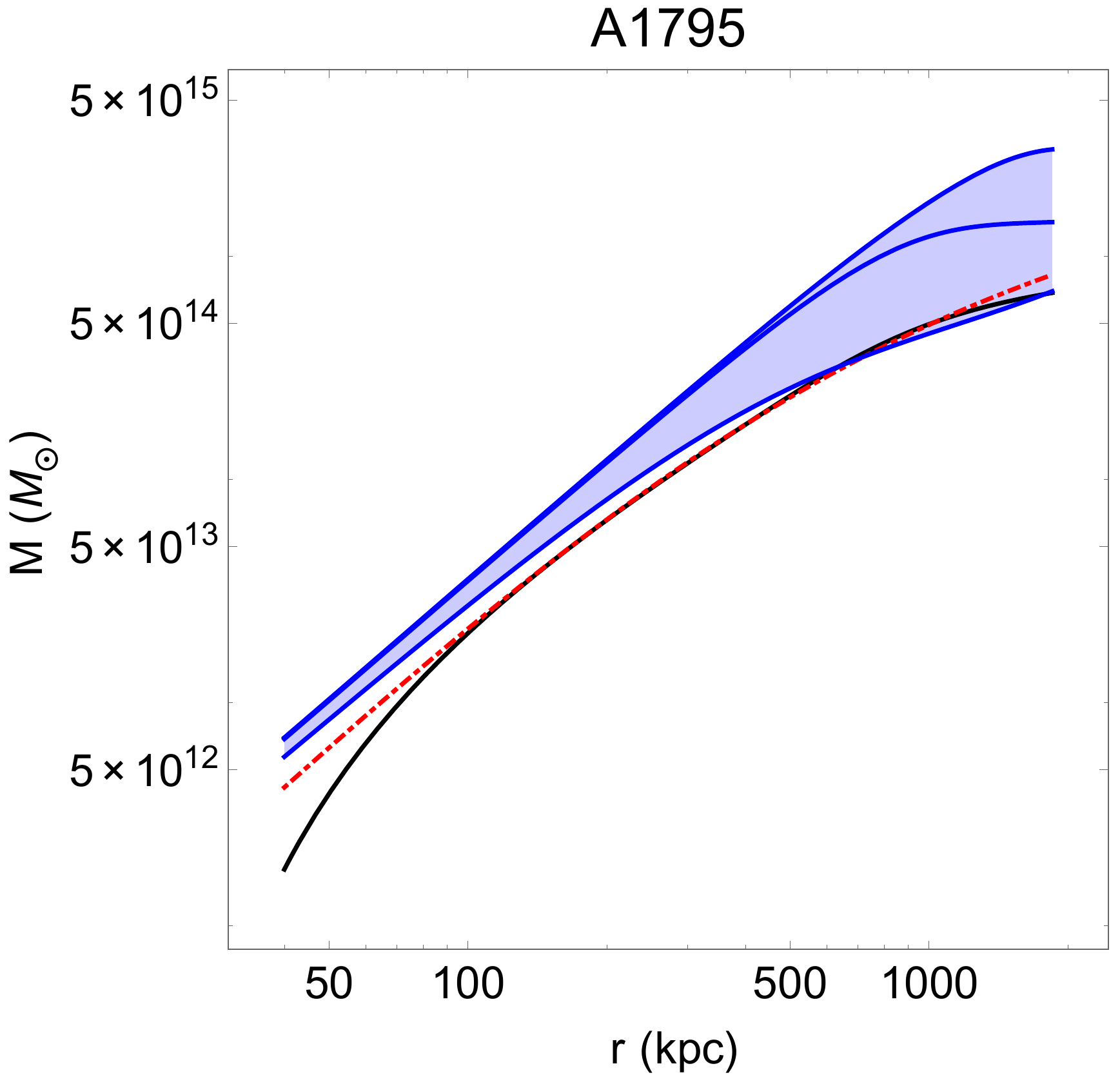} & \includegraphics[scale=0.5]{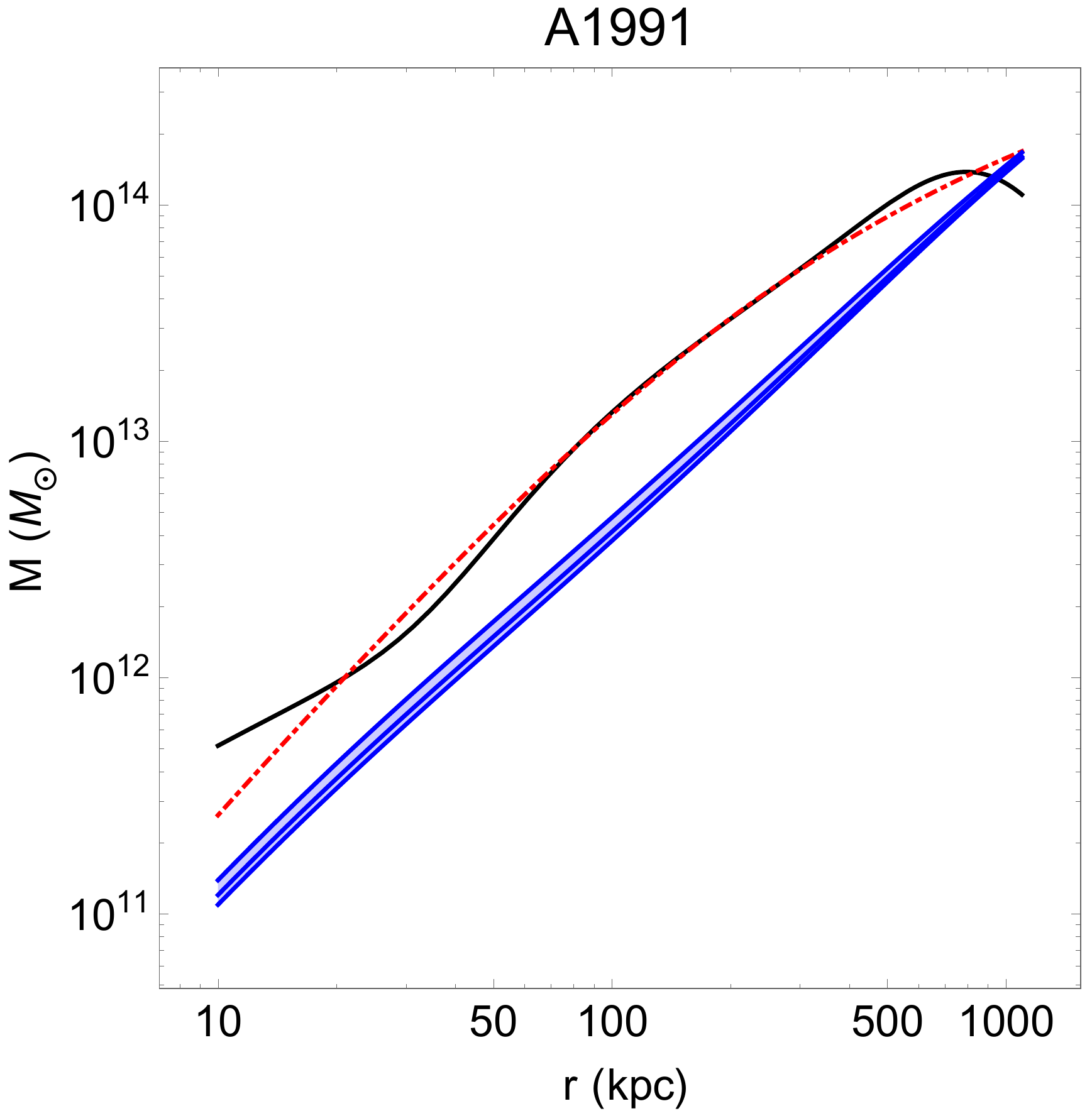}
\end{tabular}
\caption{Same as Figure \ref{MassPlotEMOND} for clusters A1795 and A1991.}
\label{}
\end{figure*}

\begin{figure*}
\begin{tabular}{ccc}
\includegraphics[scale=0.5]{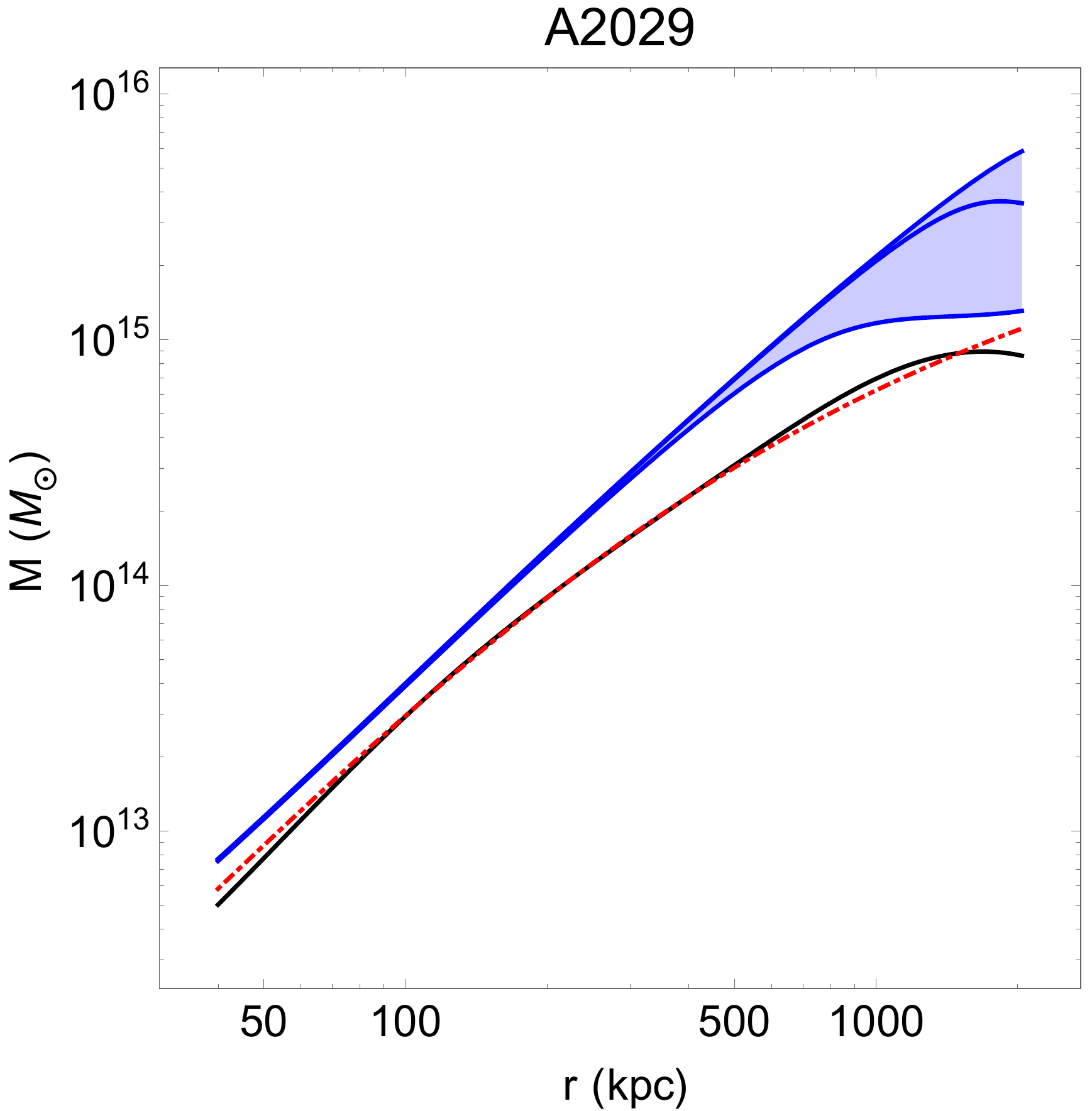} & \includegraphics[scale=0.5]{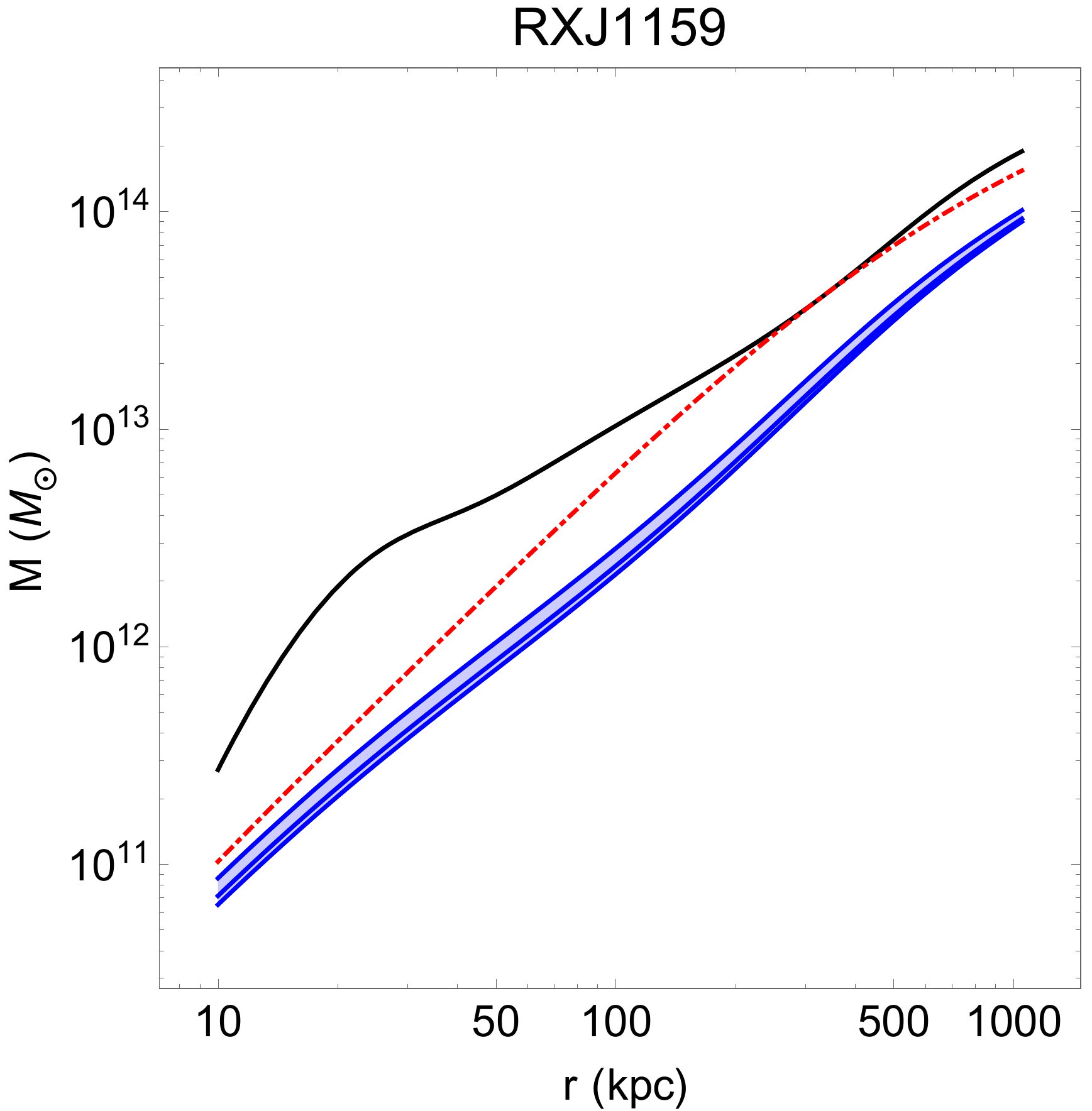}
\end{tabular}
\caption{Same as Figure \ref{MassPlotEMOND} for clusters A2029 and RXJ1159.}
\label{}
\end{figure*}

\begin{figure*}
\begin{tabular}{ccc}
\includegraphics[scale=0.5]{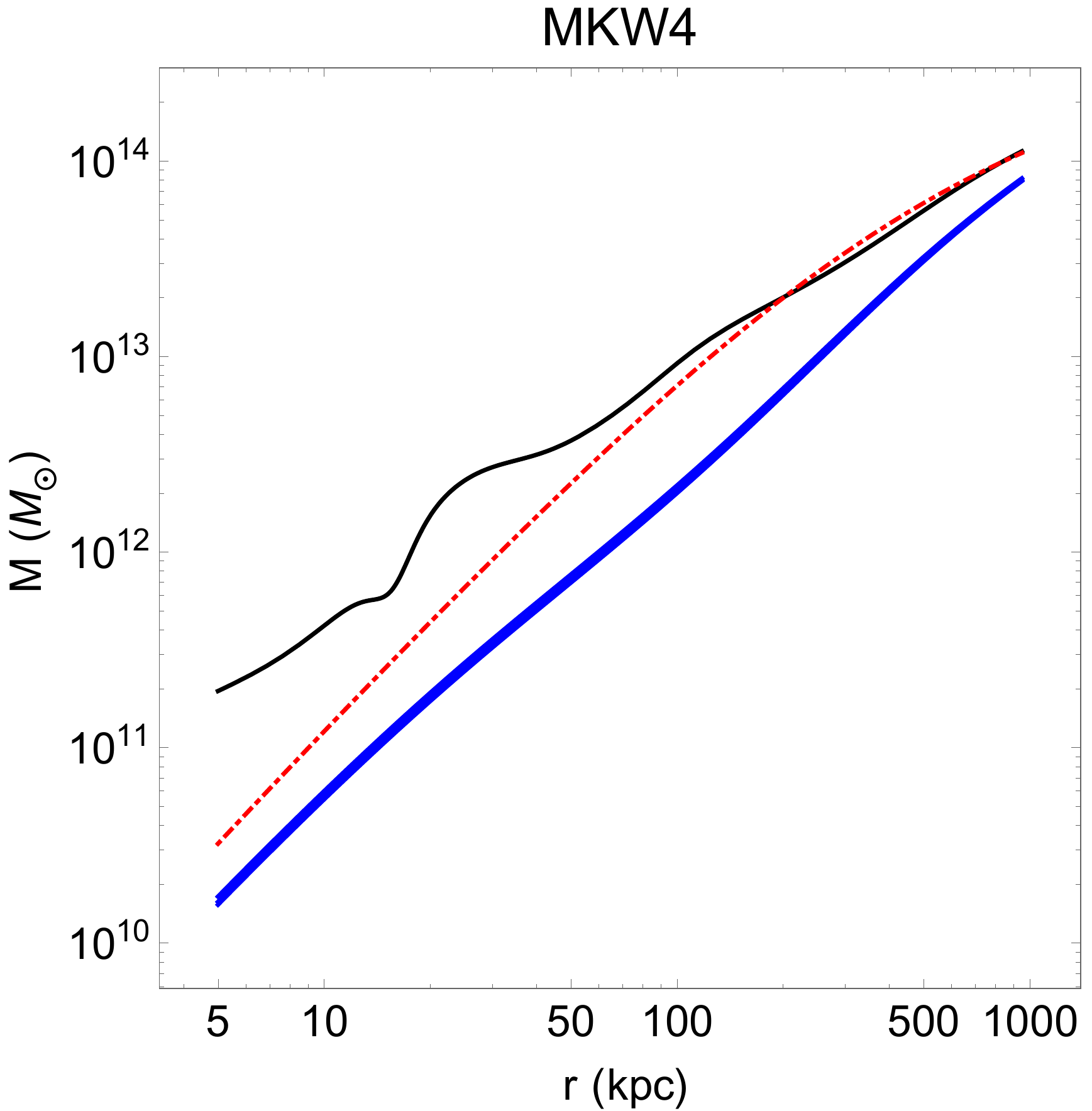} & \includegraphics[scale=0.5]{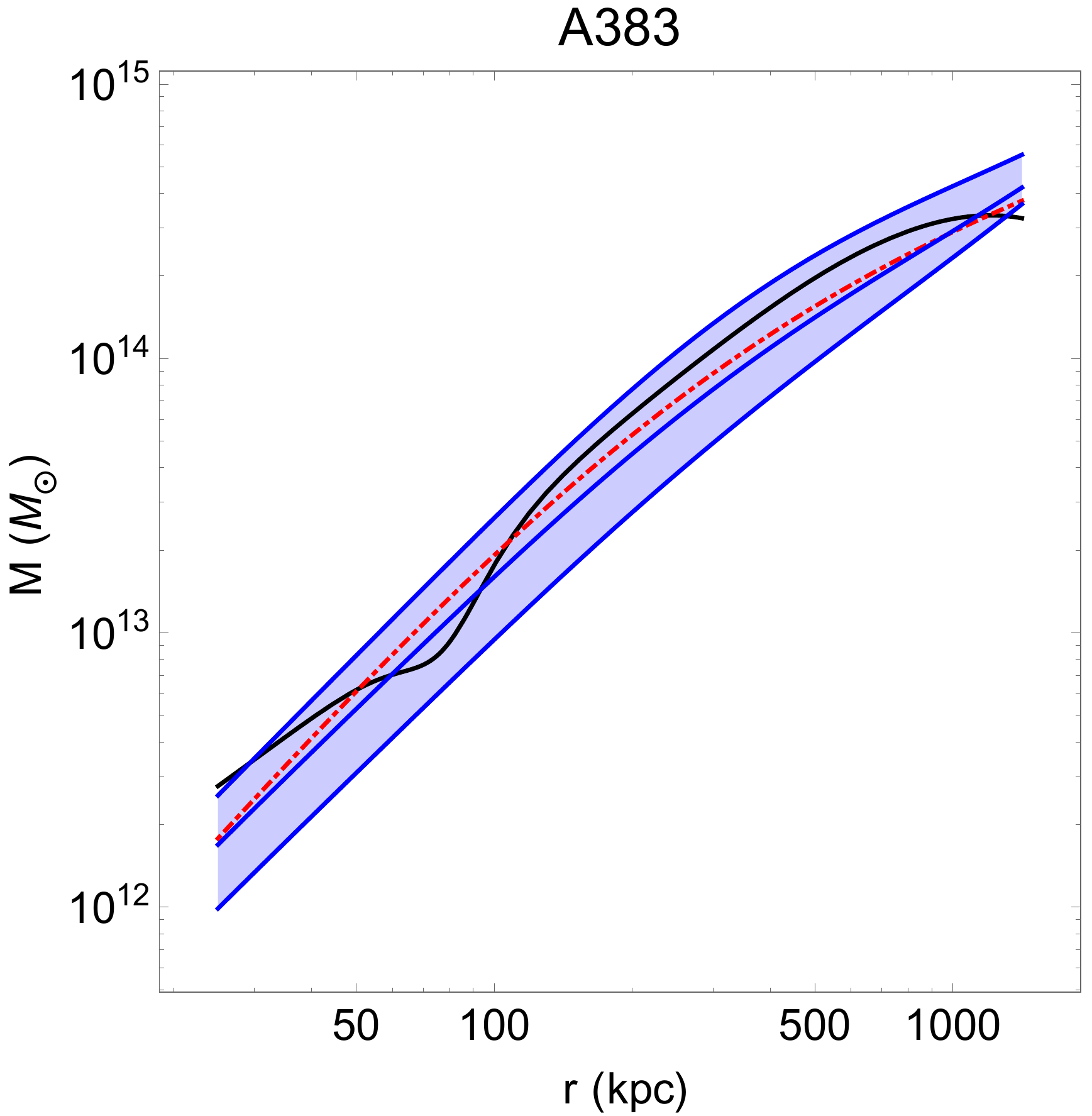}
\end{tabular}
\caption{Same as Figure \ref{MassPlotEMOND} for clusters MKW4 and A383.}
\label{}
\end{figure*}

\bibliographystyle{aa} 
\bibliography{bibEMOND} 

\end{document}